\documentclass[prb,twocolumn,amsmath,amssymb,floatfix,footinbib,bibnotes,longbibliography]{revtex4}
\pdfoutput=1

\usepackage[colorlinks=true,citecolor=blue,linkcolor=magenta]{hyperref}
\usepackage{soul}
\usepackage{multirow}
\usepackage{amsmath}
\usepackage{bbm}
\usepackage{graphicx}
\usepackage{dsfont}
\usepackage{changepage}
\usepackage{fancyhdr}
\usepackage{amsthm, amssymb}
\usepackage{array}
\newcolumntype{P}[1]{>{\centering\arraybackslash}p{#1}}
\usepackage[usenames,dvipsnames]{color}
\usepackage[section]{placeins}
\usepackage[T1]{fontenc}
\usepackage[latin9]{inputenc}
\usepackage[active]{srcltx}
\usepackage{color}
\usepackage{amssymb}
\usepackage{esint}
\usepackage[version=4]{mhchem}
\usepackage{comment}
\usepackage{natbib}
\graphicspath{ {Plots} }

\newcommand {\apgt} {\ {\raise-.5ex\hbox{$\buildrel>\over\sim$}}\ }
\newcommand {\aplt} {\ {\raise-.5ex\hbox{$\buildrel<\over\sim$}}\ }

\newcommand {\rem}[1]{}
\def  \w{\omega}
\def  \qinv{Q^{-1}}
\def  \G{\Gamma}
\def  \q0{\frac{\w_k^2}{\G}+z}

\def \titlename {Inverse Superconductor-Insulator Transition in Weakly Monitored Josephson Junction Arrays}
\def \authorf{Purnendu Das$^1$}
\def \affiliationf{$^1$Department of Physics, Indian Institute of Science, Bangalore 560012, India}
\def \authors{Sumilan Banerjee$^2$}
\def \affiliations{$^2$Centre for Condensed Matter Theory, Department of Physics, Indian Institute of Science, Bangalore 560012, India}

\begin{document}
	\title{\titlename}
	\author{\authorf, \authors}
	\affiliation{\affiliationf,\affiliations}
	\email{purnendudas@iisc.ac.in}
	\email{sumilan@iisc.ac.in}
	\date\today

\maketitle 
{\bf Control and manipulation of quantum states by measurements and bath engineering in open quantum systems have emerged as new paradigms in many-body physics. Here, taking a prototypical example of Josephson junction arrays (JJAs), we show how repetitive monitoring through continuous weak measurements and feedback can transform an insulating state in these systems to a superconductor and vice versa. We show that, even in the absence of any external thermal bath, the monitoring leads to a long-time steady state characterized by an effective `quantum' temperature in a suitably defined semiclassical limit. However, we show that the quantum dissipation due to monitoring has fundamental differences with equilibrium quantum and/or thermal dissipation in the well-studied case of JJAs in contact with an Ohmic bath. In particular, using a variational approximation, and by considering various limiting cases, we demonstrate that this difference can give rise to re-entrant steady-state phase transitions, resulting in unusual \emph{inverse} transition from an effective low-temperature insulating normal state to superconducting state at intermediate temperature. Our work emphasizes the role of quantum feedback, that acts as an additional knob to control the effective temperature of non-equilibrium steady state leading to a phase diagram, not explored in earlier works on monitored and open quantum systems.}

Measurement plays an important role at the foundation of quantum mechanics \cite{Landau1981Quantum,Wiseman:2009rda}. Measurement abruptly collapses a wave function making the evolution of quantum state non-unitary, and can disentangle an entangled state. 
The competition between entangling unitary evolution and disentangling local measurements can lead to unusual dynamical phase transitions, the so-called `measurement-induced phase transition' (MIPT) in the entanglement properties of quantum many-body systems subjected to repeated measurements \cite{Skinner2019,Choi2020,Jian2020,Li2019,Poboiko2024}. 
% MIPT is an entanglement phase transition as a function of measurement strength between a volume-law to area-law entangled long-time steady states \cite{Skinner2019,Choi2020,Jian2020,Li2019,Gullans2020,Alberton2021,Bao2020,Sang2021,Block2022,Poboiko2024,Jian2021,Nahum2021,Tang2020}. 
However, this type of measurement-induced transition is hard to realize experimentally, as the transition occurs at the level of a typical individual quantum trajectory that is characterized by keeping the full record of all the measurement outcomes. Here we study repeated measurement-induced steady-states and associated phase transitions, like quantum phase transition (QPT) between symmetry-broken and unbroken states, at the level of trajectory average quantities. For instance, we ask whether a system can be transformed from its normal state to superconducting state and vice versa, by performing repeated measurements.

To address these questions, we consider the paradigmatic system of Josephson junction arrays (JJAs) \cite{Voss1982,Schon1990,Fazio2001}, that realizes a superconductor-normal state (insulator) (SI) quantum phase transition. JJAs are physical realizations of the Quantum XY model \cite{Fazio2001,Sondhi1997}. They are artificially created as the networks of mesoscopic superconducting (SC) islands, interconnected by Josephson junctions, or may be realized as effective models for granular sueprconductors \cite{Orr1986,Haviland1989,Fazio2001}. JJAs undergo SI transition as a function of the ratio of the charging energy ($E_c$), that controls the quantum fluctuations of Cooper pair number in an island, and the Josephson coupling strength ($J$). Remarkably, dissipation, either originating intrinsically \cite{Fazio2001,Orr1986,Penttila1999} or introduced in a controlled manner \cite{Rimberg1997,Wengblast1997,Takahide2000} in experiments, is known to play important role in determining the nature of the thermodynamic SI transition. The effects of dissipation on JJAs, modeled by JJAs coupled to various types of thermal and quantum baths \cite{Fazio2001}, e.g., Ohmic baths, have been widely studied \cite{Chak1986PRL,Chak1988PRB,Kampf1987,Capriotti2005}. Possibility of dissipative phase transition even for a single Josephson junction coupled to an equilibrium bath has been studied in a large number of works \cite{Schmid1983,Bulgadaev1984,Schon1990,Werner2005} and the existence of such transition, or lack thereof, in various limits has been discussed \cite{Murani2020,Kaur2021}. In this work, we study the effects of continuous weak measurement and feedback on otherwise isolated JJAs in the absence of any external thermal bath. The measurements and feedback act as a source of non-unitary evolution and dissipation, and lead to an effective temperature description in a suitably defined classical or `high-temperature limit'. However, we show that `quantum dissipation' due to monitoring, in general, has fundamental differences with well-studied cases of equilibrium quantum and/or thermal dissipation in JJAs. We demonstrate that this difference can give rise to a steady-state phase transition from an effective low-temperature insulating normal state to superconducting state at higher temperatures, namely an \emph{inverse SI transition}.

Bath engineering to manipulate dissipation and phenomena beyond equilibrium has been discussed in the context of Lindblad description of open systems in many earlier works \cite{Sieberer2023,Diehl2008,Diehl2010a,Diehl2010b,Tomadin2011,Sieberer2013}, e.g., in the pioneering works by Diehl et al.\cite{Diehl2008,Diehl2010b} on interacting bosons with dissipative coupling to the particle current. We show that the Lindblad dynamics in this model is closely connected to the trajectory averaged dynamics in our monitored JJAs model. In the dissipative Bose-Hubbard model, a pure Bose-Einsetin condensate (BEC) in the non-interacting limit, and a weakly mixed superfluid state in the presence of interaction are induced as steady state by the engineered dissipation. In the interacting superfluid steady state, an effective temperature, controlled by the interaction strength and boson density, emerges. Our model realizes the same phenomena as a particular limit of the monitored JJA, while giving a measurement-based interpretation of the Lindblad dynamics. We show that quantum feedback in this situation acts as an additional independent handle to control the effective temperature of non-equilibrium steady state leading to a phase diagram with re-entrant and inverse superconductor-to-insulator transitions, not realized in earlier works \cite{Diehl2008,Diehl2010b,Tomadin2011}. To this end, we show that a description in terms of a single effective temperature is not adequate to obtain the inverse SI transition in the `low-temperature' or quantum limit.

\begin{figure}[ht]
\includegraphics[width=0.5\textwidth]{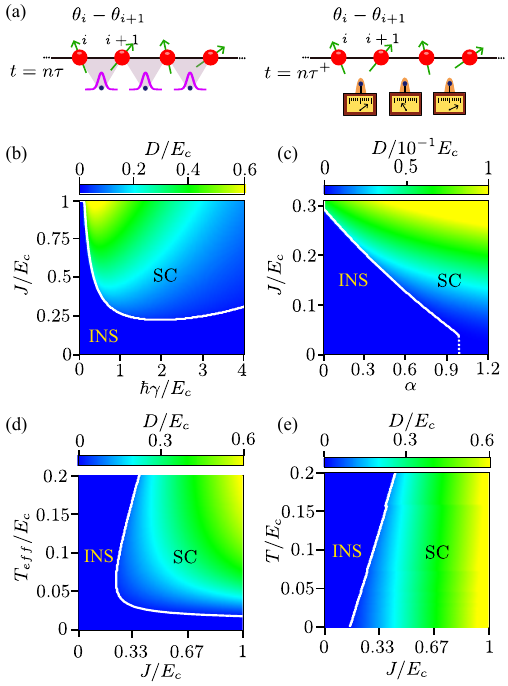}
%\centerline{\includegraphics[width=1.0\textwidth]{Fig1.pdf}}
    \caption{ {\bf Monitored Josephson junction array (JJA) and phase diagrams:} (a) The schematics of the continuous weak measurement setup for a one-dimensional (1d) JJA is shown. For the $n$-th measurement the detectors on the JJ links, e.g., $(i,i+1)$, prepared in a Gaussian state just before ($t = n\tau^-$) the measurement, are coupled to the JJ phase differences ($\hat{\theta}_i-\hat{\theta}_{i+1}$) at time $t = n\tau$. The detectors are subsequently decoupled and measured at time $t = n\tau^+$. Further, a feedback is applied using the measurement outcomes (see the main text). (b) The phase diagram within the self-consistent harmonic approximation (SCHA) as a function of JJ coupling $J$ and feedback $\gamma$ exhibits superconductor (SC)-insulator (INS) transition, as indicated by the superfluid stiffness $D$ (color) for a fixed measurement strength $\hbar\Delta^{-1}/E_c=0.5$. (c) The $J-\gamma$ phase diagram of the monitored JJA is contrasted with zero-temperature phase diagram in the $J$- dissipation ($\alpha$) plane for Ohmic JJA (see Supplementary Sec.\ref{sec:OhmicJJA_S}). (d) Phase diagram for monitored JJA as a function of effective temperature $T_{eff}$ and $J$ is compared with the $T$ vs. $J$ phase diagram of Ohmic JJA for $\alpha=0.5$ in (e).}
    \label{fig1}
\end{figure}

As depicted in Fig.\ref{fig1}(a), by generalizing a well-known model of continuous weak position measurement on a single particle \cite{Caves1987}, we construct a model of monitored JJA. In this model, the phase difference between two neighboring SC island across a JJ link is continuously and weakly measured with strength $\Delta^{-1}$ through coupling with a detector or meter, and, subsequently, a feedback \cite{Caves1987,Wiseman:2009rda} with strength $\gamma$ is applied. The measurement of phase difference plays a similar role to the dissipative current coupling in Refs.\onlinecite{Diehl2008},\onlinecite{Diehl2010b},\onlinecite{Tomadin2011}. In the presence of JJ coupling, the measurement is naively expected to enforce a fixed phase difference by weakly collapsing the wave function, and thus may favor a globally phase-coherent SC steady-state by suppressing quantum fluctuation of the SC phase when repeated many times across the JJA.
The feedback acts as a source of dissipation and maintains an effective temperature $T_{eff}\propto \Delta^{-1}/\gamma$, which diverges for $\gamma \to 0$. $T_{eff}$ is defined via an emergent fluctuation-dissipation ratio (FDR) for the long-time steady state for low-energy excitations and/or a high-temperature or classical limit. In the latter regime FDR is identical to that of the high-temperature limit of the JJAs with standard Ohmic dissipation \cite{Chak1986PRL,Chak1988PRB}. However, in the effective low-temperature or quantum limit, we find that the FDR, and thus the quantum dissipation, in monitored JJA is intrinsically different from that in the Ohmic JJA. 

Within a variational self-consistent harmonic approximation (SCHA), this difference in the nature of dissipation leads to a distinct phase diagram of monitored JJA, as a function of effective dissipation strength $\gamma$ or effective temperature, compared to that of dissipative JJA, as shown in Figs.\ref{fig1}(b-e). In particular, below a threshold value of JJ coupling $(J/E_c)_{th}$, the system remains insulating irrespective of $\gamma$. This is unlike Ohmic JJA where dissipation, above a threshold value restores superconductivity at low temperature [Fig.\ref{fig1}(c)] by inhibiting quantum phase fluctuations. Moreover, re-entrant insulator-SC-insulator transitions are realized for monitored JJA above $(J/E_c)_{th}$ [Fig.\ref{fig1}(b)]. Similarly, qualitatively different phase diagrams for monitored and Ohmic JJAs [Figs.\ref{fig1}(d,e)] are obtained as a function of effective temperature for a fixed dissipation. Again, above a threshold $(J/E_c)_{th}$, re-entrant transitions appear leading to a low-temperature insulator to intermediate-temperature SC transition. We verify our conclusions by going beyond SCHA, in the (i) high-temperature or classical, (ii) strong measurement, (iii) strong feedback, and (iv) weak JJ coupling limits. We show that the behavior in large feedback limit, and the renormalization group (RG) flow for the weak-coupling limit of the monitored JJA are different from those in Ohmic JJA, in conformity with the SCHA. On the contrary, a stochastic Langevin description emerges in the monitored system in the classical limit at high temperatures or strong measurements. This leads to a classical FDR controlled by $\gamma$ and $T_{eff}$. As a result, the SI transition in this limit is identical to finite-temperature SI transition in Ohmic JJA.

\noindent{\bf Model of monitored JJA}

We consider a system-detector model [Fig.\ref{fig1}(a)] with Hamiltonian $\mathcal{H}(t)=\mathcal{H}_s+\mathcal{H}_{sd}(t)$, where the system of JJA \cite{Fazio2001} is described by $\mathcal{H}_s=(1/2)\sum_i E_c \hat{n}_i^2+J\sum_{\langle ij\rangle}(1-\cos{(\hat{\theta}_i-\hat{\theta}_j)})$, with on-site charging energy $E_c$ and nearest-neighbor JJ coupling $J$ on a hypercubic lattice in $d$ dimension with $N$ sites. The Cooper pair number $\hat{n}_i$ and the phase $\hat{\theta}_i$ operators on each SC island $i$ satisfies $[\hat{\theta}_i,\hat{n}_j]=\mathrm{i}\delta_{ij}$. The JJA is coupled via $\mathcal{H}_{sd}(t)=\hbar\sum_{\langle ij\rangle }\delta(t-n\tau)\Delta\hat{\theta}_{ij}\hat{p}_{ij,n}$ to a set of detectors or `meters' at regular intervals $t=n\tau$ through the phase difference $\Delta\hat{\theta}_{ij}=\hat{\theta}_i-\hat{\theta}_j$ across JJ at the link $\langle ij\rangle $. The meters are quantum particles with \emph{dimensionless} position $\hat{x}_{ij,n}$ and momenta $\hat{p}_{ij,n}$ satisfying $[\hat{x}_{ij},\hat{p}_{kl}]=\mathrm{i}\delta_{\langle ij\rangle,\langle kl\rangle}$, where the strength of the delta function system-detector coupling has been absorbed in defining the unit of the detector position and momentum. The meters are prepared in a Gaussian state $\psi_{ij}(x_{ij,n})=(\pi\sigma)^{-1/4}\exp{(-x_{ij,n}^2/2\sigma)}$ before the measurement at $t=n\tau^-$. The weak measurements on the system comprise projective measurements of the position of the meters at $t=n\tau^+$ leading to stochastic outcomes $\xi_{ij,n}$ at the $n$-th set of measurements on all the links. The full record of all the measurement outcomes $\{\xi_{ij,n}\}$ over a time interval constitutes a quantum trajectory. However, these repeated measurements lead to large fluctuations of Cooper pair number at a site and the systems heats up, as known for the case of weak position measurements of a single particle \cite{Caves1987}. Hence, we apply a unitary feedback $U_{ij,n}=e^{\mathrm{i}\gamma \tau \xi_{ij,n}(\hat{n}_i-\hat{n}_j)}$ after each measurement with strength $\gamma$ to introduce a source of dissipation \cite{Caves1987,Wiseman:2009rda}.

\begin{figure}[ht]
\includegraphics[width=0.5\textwidth]{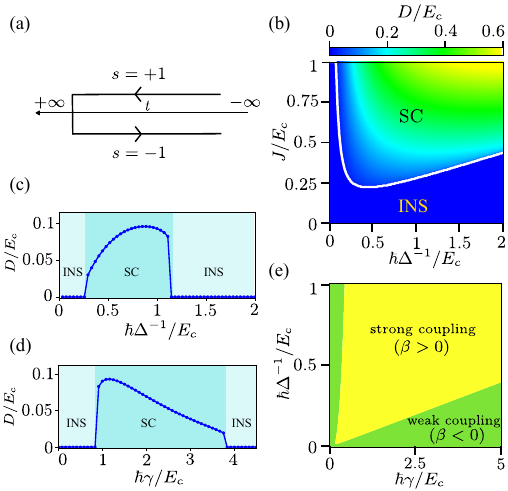}
    \caption{{\bf Phase diagram of monitored JJA and weak-coupling renormalization group:} (a) Schwinger-Keldysh contour, with forward ($s=+1$) and backward ($s=-1$) branches, used to describe non-unitary dynamics in the long-time steady of monitored JJA. (b) Phase diagram as a function of JJ coupling $J$ and measurement strength $\Delta^{-1}$ for a fixed feedback $\hbar\gamma/E_c=2$. The color indicates the superfluid stiffness $D$ that demarcates between insulating (INS) and superconducting (SC) states. The re-entrant insulator-superconductor-insulator transitions are shown in $D$ vs. $\Delta^{-1}$ plot in (c) for $J/E_c=0.3$ and $\hbar\gamma/E_c=2$, and in $D$ vs. $\gamma$ plot in (d) for $J/E_c=0.3$ and $\hbar\Delta^{-1}/E_c=0.5$. (e) The weak-JJ coupling renormalization group qualitatively agrees with the re-entrant nature of the phase diagrams as a function of $\gamma$ and $\Delta^{-1}$. The sign of the $\beta$ function (color) indicates the flow towards weak coupling ($\beta<0$, insulator) or strong coupling ($\beta>0$, superconductor).}
        \label{fig2}
\end{figure}

As we show in the Methods, in the above measurement model, one can obtain a non-unitary time evolution of the (unnormalized) density matrix $\rho(t)$ of the system. Moreover, in the limit of continuous weak measurement limit \cite{Caves1987}, $\tau\to 0$, $\sigma\to \infty$ such that $\sigma\tau=\Delta$ is finite, the dynamics of the JJA system for a particular quantum trajectory $\{\xi_{ij}(t)\}$ over a time interval $t_f$ can be described by a Schwinger-Keldysh (SK) [Fig.\ref{fig2}(a)] generating function $Z[\xi]=\mathrm{Tr}[\rho(t_f,\{\xi\})]=\int \mathcal{D}\theta e^{\mathrm{i}S[\theta,\xi]/\hbar}\langle \theta_+(t_0)|\rho_0|\theta_-(t_0)\rangle$, which is essentially the Born probability of the trajectory. Here $\Delta^{-1}$ is the measurement strength and $\rho_0$ is the density matrix of the system at the initial time $t_0$. We assume that the non-equilibrium steady state for $t_f\to \infty $ does not depend on $\rho_0$. The SK action is given by
\begin{align}
&S[\theta,\xi]=\int_{t_0}^{t_f}dt[\sum_{i,s}s\{\frac{\hbar^2\dot{\theta}^2_{is}}{2E_c}+\frac{\hbar^2\gamma}{E_c}\dot{\theta}_{is}\sum_{j\in \mathrm{nn}_i} \xi_{ij}(t)\}\nonumber \\
&-\sum_{\langle ij\rangle,s}s\{J\left(1-\cos\Delta\theta_{ij,s}\right)-\frac{\mathrm{i}s\hbar}{2\Delta}\left(\xi_{ij}(t)-\Delta\theta_{ij,s})^2\right)\}]. \label{eq:Action}
\end{align}
Here $s=\pm (\pm 1)$ denotes the forward and backward branches of SK contour [Fig.\ref{fig2}(a)], and $\dot{\theta}_{is}=d\theta_{is}/dt$ and $\mathrm{nn}_i$ refers to sites nearest neighbor to $i$. For simplicity, we have assumed the charges $\hat{n}_i$ on the SC islands to be continuous and thus $-\infty<\theta_{is}(t)<\infty$ in the SK path integral, as is usually done for the dissipative JJA models \cite{Fazio2001}. The model can be extended to incorporate discrete nature of the charges \cite{Fazio2001}. Moreover, the total Cooper pair number $\sum_i \hat{n}_i$ is conserved under the monitored dynamics and thus fixed by the initial state. Assuming a homogeneous steady state, this can be implicitly incorporated in the model by defining $\hat{n}_i$ with respect to the mean occupation $\langle\hat{n}\rangle=(1/N)\sum_i\langle n_i\rangle$, as we do here. With this redefinition, $\sum_i \langle \hat{n}_i\rangle =0$ in the steady state and effectively $-\infty<n_i<\infty$ in the thermodynamic limit.

The dynamics of the monitored JJA, as described above through a SK path integral, can also be represented thrugh a stochastic Schr\"{o}dingr equation \cite{Gisin1984,Ghirardi1990} (see Supplementary Sec.\ref{sec:Lindblad_S}). Moreover, as we only consider the trajectory-averaged steady-state in this work, the dynamics can also be described in terms of a Lindblad master equation \cite{Breuer2002} with the effects of feedback properly included, as described in Supplementary Sec.\ref{sec:Lindblad_S}. There, we also discuss the generalization of the Lindblad dynamics with feedback for the JJA to the related dissipative Bose-Hubbard model \cite{Diehl2008,Diehl2010b}. We show that model considered by Diehl et al.\cite{Diehl2008,Diehl2010b} arises in the limit $(\Delta^{-1}/\gamma)=2\langle \hat n\rangle$ leading to $T_{eff}=E_c\langle \hat{n}\rangle/2$. However, this description in terms of only an effective temperature is not sufficient to describe the system in the quantum limit, where the full FDR needs to be taken into account to capture the inverse SI transition.

\noindent{\bf Self-consistent harmonic approximation (SCHA)}

We first use a self-consistent harmonic approximation (SCHA) where a variational action $S_{v}$ is obtained by replacing JJ coupling term $J(1-\cos\Delta\theta_{ij,s})$ in Eq.\eqref{eq:Action} with a harmonic term $D_{ij,s}(t)\Delta\theta_{ij,s}^2/2$ having the variational parameter $D_{ij,s}(t)$ (see Methods). The variational Born probability $Z_{v}[\xi]=\exp{(-F_{v}[\xi])}$, which is bounded above by the actual Born probability $Z[\xi]$, is maximized by minimizing $F_{v}=-(\mathrm{i}/\hbar)\langle (S-S_{v})\rangle_{v}-\ln Z_{v}$ with respect to $D_{ij,s}(t)$.
As discussed in the Methods, we further assume a spatio-temporal translationally invariant steady state, and solve the self-consistency condition after averaging over all the trajectories, i.e.,
\begin{align}
D= J\exp{(-\overline{\langle \Delta\theta_{ij,s}^2(t)\rangle_v}/2)}, \label{eq:SelfConsistency}
\end{align}
where $\overline{\langle \cdots\rangle_v}$ denotes trajectory averaged expectation value. The self-consistent parameter $D$ gives the superfluid stiffness and demarcates between superconductor ($D\neq 0$) and normal state ($D=0$), e.g., the insulator. 

\noindent{\bf Steady-state FDR and effective temperature}

As discussed in the Methods, the effective self-consistent harmonic action for the trajectory-averaged steady state can be written as 
\begin{align}
\frac{1}{\hbar}S_{eff}=\frac{1}{2}\sum_q \theta^T(-q)G^{-1}(q)\theta(q) \label{eq:Seff}  
\end{align}
 in terms of Fourier transform $\theta^T(q)=(\theta_c(q),\theta_q(q))$ of the classical and quantum components $\theta_c$ and $\theta_q$, after Keldysh rotation \cite{kamenev2011} $\theta_{is}(t)=\theta_{ic}(t)+s\theta_{iq}(t)$; $q=(\boldsymbol{q},\omega)$ denotes momentum and frequency, and $\sum_q\equiv \sum_{\boldsymbol{q}}\int_{-\infty}^{\infty}(d\omega/2\pi)$. The effective inverse propagator $G^{-1}(q)$ has the usual causal Keldysh structure \cite{kamenev2011}, namely $[G^{-1}]_{cc}(q)=0$, the retarded (advanced) components $[G^{-1}]_{qc(cq)}=[G^{-1}]^{R(A)}(q)=(2/\hbar)[(\hbar \omega^{\pm})^2/E_c-(D\mp\mathrm{i}\hbar^2\gamma\omega^{\pm}/E_c)K(\boldsymbol{q})]$, and $[G^{-1}]_{qq}(q)=[G^{-1}]^K(q)=(2\mathrm{i}/\hbar)K(\boldsymbol{q})(\hbar\Delta^{-1}+\hbar^3\gamma^2\omega^2/ E_c^2\Delta^{-1})$. Here $K(\boldsymbol{q})=2d-2\sum_{\mu=1}^d\cos{(q_\mu a)}$ and $\omega^\pm=\omega\pm \mathrm{i}\eta$ with $\eta\to 0^+$. The propagator $G(q)$ can be used to compute $\overline{\langle \Delta\theta_{ij,s}^2(t)\rangle_v}=(E_c/4dD)(\Delta^{-1}/\gamma)+(\gamma/2\Delta^{-1})$, and thus to obtain $D$ self consistently from Eq.\eqref{eq:SelfConsistency} (see Methods). 

Using $G(q)$, we can obtain a FDR for the steady state from $G^K(q)/[G^R(q)-G^A(q)]=2T_{eff}/\hbar\omega+\hbar \omega/8T_{eff}$, where
\begin{align}
T_{eff}&=\frac{\Delta^{-1}}{4\gamma}E_c \label{eq:EffectiveTemp}
\end{align}
is an effective temperature that becomes unbounded in the absence of the feedback. Moreover, within SCHA, the phase fluctuation can be written solely in terms of the combination $(\Delta^{-1}/\gamma)\propto T_{eff}$, i.e. $\overline{\langle \Delta\theta_{ij,s}^2(t)\rangle_v}=(T_{eff}/dD)+(E_c/8T_{eff})$.
The steady-state FDR can be compared with usual FDR for thermal equilibrium at temperature $T$, namely $\coth(\hbar\omega/2T)=2T/\hbar\omega+\hbar\omega/6T+\cdots$ ($k_\mathrm{B}=1$), which is applicable to Ohmic JJA \cite{Chak1986PRL,Chak1988PRB} (Supplementary Sec.\ref{sec:OhmicJJA_S}). Thus the steady-state FDR coincides with the thermal FDR with $T=T_{eff}$, for an effective high-temperature limit $T_{eff}\gg \hbar \omega$, or equivalently, for low energy ($\omega\to 0$) or the semiclassical limit ($\hbar\to 0$). However, unlike the Ohmic JJA \cite{Chak1986PRL,Chak1988PRB}, the temperature in the monitored JJA emerges in the absence of any thermal bath, purely from measurement, feedback and quantum fluctuations \cite{Ruidas2024}. Moreover, the steady-state FDR is completely different from the thermal FDR at low effective temperatures, leading to fundamentally different quantum dissipation and qualitatively distinct phase diagram of Fig.\ref{fig1} for monitored JJA, compared to Ohmic JJA. In particular, there is no analogous \emph{zero-temperature} limit for the monitored model compared to the Ohmic JJA, as discussed in Supplementary Sec.\ref{sec:MonitoredOhmicComp_S}. For example, the thermal FDR, $\cot{(\hbar\omega/2T)}$, approaches $\mathrm{sgn}(\omega)$ as $T\to 0$, whereas the FDR in monitored JJA diverges as $\hbar \omega/T_{eff}$ as $T_{eff}\to 0$. In Supplementary Sec.\ref{sec:OhmicJJA_SCHA_S}, comparing the propagator $G(q)$ of the monitored and Ohmic JJAs, we draw parallels between feedback $\gamma$ and Ohmic dissipation in the high-temperature limit.

\noindent{\bf Phase diagram}

We obtain the phase diagrams for the monitored JJA by numerically solving the self-consistency equation of Eq.\eqref{eq:SelfConsistency}. As shown in Figs.\ref{fig1}(b-d), the phase diagrams of monitored JJA as a function of $\gamma$ and $T_{eff}$, are compared with the phase diagrams of Ohmic JJA as a function of dissipation strength $\alpha$ and temperature $T$, respectively. For completeness, we also plot the phase diagram as a function of measurement strength $\Delta^{-1}$ for a fixed $\gamma$ in Fig.\ref{fig2}(b), and demonstrate the re-entrant transitions for a fixed $J$ in Figs.\ref{fig2}(c,d). As discussed in the Supplementary Sec.\ref{sec:PhaseDiagramAnalytic_S}, the phase boundary between SC and insulating phases within SCHA can be shown to be always first order with a discontinuity in $D$. However, the first order nature could be due to the well-known artifact \cite{Chakraverty1979,SIMANEK1980} of the SCHA near a transition. Nevertheless, the basic structure of the phase diagrams in Figs.\ref{fig1},\ref{fig2}, namely the existence of the phases and the re-entrant transitions are expected to be robust features, beyond the SCHA.

\noindent{\bf Semi-classical limit of monitored JJA}

We can obtain a non-trivial classical or semiclassical limit $\hbar\to 0$ \cite{Ruidas2024}, keeping the ratio $\Delta/\hbar^2$ fixed, as discussed in the Supplementary Sec.\ref{sec:SemiClassical_S}. This is achieved by rewriting the action in Eq.\eqref{eq:Action}, in terms $\theta_{ic}$ and $\theta_{iq}$, and expanding up to quadratic order in quantum component $\theta_{iq}$. In this limit all the quantum trajectories tend to a single trajectory with measurement outcomes $\xi_{ij}(t)=\Delta\theta_{ij,c}(t)$, i.e., $\xi_{ij}(t)$ is pinned to the classical component of the phase differences on $\langle ij\rangle$. The dynamics of the phases $\theta_{ic}(t)\equiv \theta_i(t)$ follows standard stochastic Langevin equation,
\begin{align}
\frac{\hbar^2}{E_c}\ddot{\theta}_i=\sum_{j\in \mathrm{nn}_i}[-\frac{\hbar^2\gamma}{E_c}\Delta\dot{\theta}_{ij}-J\sin\Delta\theta_{ij}+\eta_{ij}] \label{eq:LangevinEq}
\end{align}
where $\eta_{ij}$ is a Gaussian white noise originating from fluctuations of $\theta_{iq}$ with $\langle\eta_{ij}(t)\eta_{kl}(t')\rangle=(\hbar^2/2\Delta)\delta_{\langle ij\rangle,\langle kl\rangle}\delta(t-t')$. The latter, along with the damping term in
Eq.\eqref{eq:LangevinEq}, constitute the classical FDR, $2(\hbar^2/E_c)\gamma T_{eff}=\hbar^2/2\Delta$, resulting in the same effective temperature as in Eq.\eqref{eq:EffectiveTemp}. For a meaningful classical limit with non-trivial dynamics, one needs to take $E_c\propto \hbar^2$ with $\gamma\sim \mathcal{O}(\hbar^0)$, as evident from Eq.\eqref{eq:LangevinEq}. As a result, $\overline{\langle \Delta\theta^2_{ij,s}\rangle_v}=T_{eff}/dD+\mathcal{O}(\hbar^2)$ in Eq.\eqref{eq:SelfConsistency} and $D=J\exp{(-T_{eff}/2dD)}$. The latter exactly matches with the SCHA self-consistency equation of Ohmic JJA in the high-temperature classical limit (Supplementary Sec.\ref{sec:SemiClassical_S}), implying that the monitored and Ohmic JJAs behave identically in this limit. In fact, the Langevin equation [Eq.\eqref{eq:LangevinEq}] can be used to study this effectively thermal SI transition beyond SCHA, and it will lead to a Berezinskii-Kosterlitz-Thouless (BKT) transition \cite{Berezinsky:1970fr,JMKosterlitz_1973} in $d=2$ and a continuous $XY$ transition in $d=3$.

\noindent{\bf Strong measurement limit of monitored JJA}

As discussed above, within SCHA and in the semiclassical limit, $T_{eff}\propto \Delta^{-1}$. Hence, we expect that the strong measurement limit $\hbar\Delta^{-1}\gg \hbar \gamma,J, E_c$ also leads to an effective \emph{semiclassical limit} for $\hbar\neq 0$. As we show in Supplementary Sec.\ref{sec:StrongMeasurement_S}, the dynamics of the monitored JJA indeed becomes classical in the strong measurement limit and is described by the same Langevin dynamics of Eq.\eqref{eq:LangevinEq}. Thus, strong measurements of the phase difference across JJ links destroy SC state, even though one naively expects to such measurements to act in synergy with JJ coupling by fixing the phase differences. However, in the strong measurement limit $\Delta\to 0$, the phase differences are collapsed completely randomly and the measurements strength acts like an effective randomizing temperature against the tendency of global phase coherence due to the JJ coupling. 

\noindent{\bf Strong feedback limit of monitored JJA}

The feedback $\gamma$ appears analogous to the dissipation $\alpha$ in the Ohmic model, when retarded/advanced propagator and high-temperature limits within SCHA are compared between the two cases (Supplementary Sec.\ref{sec:OhmicJJA_SCHA_S}). However, as shown in Figs.\ref{fig1}(b) and (c), the SC state is destroyed for large feedback strength $\gamma$, unlike the Ohmic dissipation. This can be understood by going beyond the SCHA in the strong feedback limit $\hbar\gamma \gg J,E_c$ limit (Supplementary Sec.\ref{sec:StrongFeedbackRG_S}). In this case, one finds the fate of the SC state by looking at renormalized JJ coupling of the slow phase modes after integrating out the fast modes. To this end, we divide the phase variables into fast ($\tilde{\omega}_c=JE_c/\hbar^2\gamma \lesssim\omega\lesssim \omega_c=\gamma$) and slow ($0<\omega\lesssim \omega_c$) modes, namely, $\theta^>_{is}(t)=\int_{\tilde{\omega}_c\leq |\omega|\leq \omega_c} (d\omega/2\pi)e^{-\mathrm{i}\omega t}\theta_{is}(\omega)$ and $\theta^<_{is}(t)=\int_{-\omega_c}^{\omega_c}(d\omega/2\pi)e^{-\mathrm{i}\omega t}\theta_{is}^<(\omega)$. Treating $J$ perturbatively while integrating the fast modes, this procedure leads to an action for the slow modes with the effective coupling $J_{eff}=J(1-\overline{\langle (\Delta\theta_{ij,c}^>)^2\rangle_>}/2)$, where $\overline{\langle (\Delta\theta_{ij,c}^>)^2 \rangle_>}$ is evaluated for the fast modes (see Supplementary Sec.\ref{sec:StrongFeedbackRG_S}). Considering the limit $\hbar\gamma\gg \hbar \Delta^{-1},J,E_c$, we obtain $J_{eff}\simeq J (1-\gamma/4\Delta^{-1})$, and $J_{eff}\simeq J(1-(\Delta^{-1}/4\pi\gamma)(E_c/Jd))$ for $\hbar \Delta^{-1}\gg \hbar \gamma\gg J,E_c$. This implies that $J_{eff}$ becomes negative in these strong feedback limits, indicating the absence of SC steady state, consistent with the SCHA phase diagram in Figs.\ref{fig1}(b),\ref{fig2}(b). The absence of SC phase in monitored JJA in the strong feedback limit is completely opposite to the case of the strong dissipation limit of Ohmic JJA. In the latter, for large $\alpha$, $J_{eff}\simeq J (1-(1/2\alpha d)\ln[1+(\alpha/2\pi)])\to J$, implying a SC phase for $\alpha\gg 1$ \cite{Chakraverty1979}. This difference in quantum dissipation between the monitored and Ohmic cases can be understood from the FDR, which diverges for $T_{eff}\propto \gamma^{-1} \to 0$ in the monitored case and remains finite in the Ohmic JJA. As a result strong quantum fluctuations destroy the SC state in the strong feedback limit in the monitored case.

\noindent {\bf Weak JJ coupling limit of monitored JJA}

We can infer the fate of SC phase through the perturbative RG flow equation for the JJ coupling $J$ in the weak-coupling limit $J\ll E_c$. We again proceed by defining fast, $\theta_{is}^>(t)=\int_{\omega_c/b\leq |\omega|\leq \omega_c}(d\omega/2\pi)e^{-\mathrm{i}\omega t}\theta_{is}(\omega)$, and slow, $\theta_{is}^<(t)=\int_{-\omega_c}^{\omega_c}(d\omega/2\pi)e^{-\mathrm{i}\omega t}\theta_{is}(\omega)$, phase modes. Here $\omega_c$ is the cutoff frequency, determined by the largest of $E_c$, $\hbar \gamma$, and $\hbar \Delta^{-1}$. As discussed in the Supplementary Sec.\ref{sec:StrongFeedbackRG_S}, the cutoff is systematically rescaled by $b=e^{\delta l}$ to integrate out the fast modes perturbatively up to first order in $J/E_c$ for the effective action obtained from trajectory averaged generating function $\mathcal{Z}=\int \mathcal{D}\xi Z[\xi]$. 
% After rescaling $(\theta_{is}(\omega)/b,\omega b,t/b, E_c b)\to (\theta_{is}(\omega),\omega,t,E_c)$, to restore the original form of the action and the original cutoff, 
Through usual RG steps, this results in a renormalized JJ coupling $J(b)=Jbe^{-\langle (\Delta\theta_{ij,s}^{>})^2\rangle_{>}/2}$. Performing infinitesimal RG transformation $b=e^{\delta l}\approx 1+\delta l$, we obtain the RG flow equation, $d\ln{J}/d\ln{b}=\beta(\tilde{\gamma},\tilde{\Delta}^{-1})$, where $\beta(\tilde{\gamma},\tilde{\Delta}^{-1})=[1-(1/4\pi \tilde{\gamma}d)(\tilde{\Delta}^{-1}/\tilde{\gamma}^2+(\tilde{\gamma}^2/\tilde{\Delta}^{-1}))]$ with $(\tilde{\gamma},\tilde{\Delta}^{-1})=(\hbar\gamma/E_c,\hbar\Delta^{-1}/E_c)$. As shown in Fig.\ref{fig2}(e), for $\beta(\tilde{\gamma},\tilde{\Delta}^{-1})>0$, the JJ coupling flows to strong coupling, presumably, suggesting a SC state, whereas $\beta(\tilde{\gamma},\tilde{\Delta}^{-1})<0$ implies that $J$ flows to zero, and indicates an insulating state. Fig.\ref{fig2}(e) is consistent with the re-entrant transitions in the SCHA phase diagram [Fig.\ref{fig1}(b,d), \ref{fig2}(b)]. This can be contrasted with the case of Ohmic JJA \cite{Chak1986PRL,Chak1988PRB}, where $d\ln{J}/d\ln{b}=\beta(\alpha)=(1-1/\alpha d)$ indicating the flow to strong coupling (SC state) for dissipation $\alpha>1/d$ and $J\to 0$ for $\alpha<1/d$. In this work, we do not study the flow of the FDR which may change at higher order in $J$, unlike in the Ohmic JJA where FDR remains the same order by order in $J$ due to fluctuation dissipation theorem at thermal equilibrium. The flow of FDR may have important consequences for the critical properties of the SI transition at intermediate effective temperatures in monitored JJA. 

% \noindent {\bf Conclusions}\\
% In this work, we have explored the phase diagram of long-time steady state of weakly and continuously measured JJA with feedback control by looking into quantities, e.g., superfluid density, averaged over quantum trajectories. To this end, we develop a variational self-consistent harmonic approximation based on Born probabilities of quantum trajectories within a Schwinger-Keldysh field theory for non-unitary dynamics in the long-time steady state.  
% Our work reveals an intriguing insulator-superconductor-insulator re-entrant phase transition as a function of the measurement strength or feedback. We find that the monitored model behaves exactly like the JJA with Ohmic dissipation under certain conditions, such as high effective temperatures or semi-classical limit. However, at low effective temperatures, the monitored JJA fundamentally differs from the dissipative model. For example, there are no analogous limits for the monitored model when compared to the zero-temperature Ohmic model. Our work highlights crucial role of feedback to unravel unusual dynamical phase transitions in non-equilibrium steady-state, previously unexplored in engineered open quantum systems, and without any analogue in equilibrium systems, such as low-temperature insulator to high-temperature superconductor transition.

% Our work here highlights crucial role of feedback to unravel unusual dynamical phase transitions in non-equilibrium steady-state, without any analogue in equilibrium systems. 
In future, our explicit system-detector model with measurement and feedback can be used to theoretically study the measurement-induced entanglement transition at the level of typical individual trajectory for pure state evolution, or even trajectory-averaged mixed-state transitions in terms of mixed-state entanglement measures such as entanglement negativity. Our model of monitored JJA can be directly realized in superconducting-qubit based systems \cite{Makhlin2001,Kjaergaard2020}, which are of great current interest as one of the most promising platforms for quantum computing. Both strong and weak measurements, as well as feedback, can be realized in these systems, as has been demonstrated \cite{Korotkov2001,Sayrin2011,Vijay2012,Murch2013,Weber2014}. To realize our monitored JJA, the current across JJ links, for example, can be weakly measured by inductively coupling it to current through linear inductors and by coupling the inductor to a microwave transmission line \cite{Vool2016}. The feedback involving charges in the SC island can be applied capcaitively based on the measurement outcomes. Such realizations can pave the way to understand the role of fluctuations and dissipation due to \emph{measurement or observer baths}, as opposed to usual thermal bath, and thus to control and manipulate many-body states in a quantum computer.

\section*{Acknowldegements}
We acknowledge fruitful discussions with Vibhor Singh, Benjamin Huard, David Logan and Abhishek Dhar. PD acknowledges support from Kishore Vaigyanik Protsahan Yojana (KVPY) from DST, Govt. of India. SB acknowledges support from CRG, SERB (ANRF), DST, India (File No. CRG/2022/001062) and STARS, MoE, Govt. of India (File. No. MoE-STARS/STARS-2/2023-0716).

%\appendix

%%%%%%%%%%%%%%%%%%%%%%%%%%%%%%%%%%%%%%%%%%%%%%%%%%%
%%%%%%%%%%%%%%%%%%%%%%%ta%%%%%%%%%%%%%%%%%%%%%%%%%%%%
%%%%%%%%%%%%%%%%%%%%%%%%%%%%%%%%%%%%%%%%%%%%%%%%%%%
%%%%%%%%%%%%%%%%%%%%%%%%%%%%%%%%%%%%%%%%%%%%%%%%%%%
%%%%%%%%%%%%%%%%%%%%%%%%%%%%%%%%%%%%%%%%%%%%%%%%%%%

%%%%%%%%%%%%%%%%%%%%%%%%%%%%%%%%%%%%%%%%%%%%%%%%%%%
%%%%%%%%%%%%%%%%%%%%%%%%%%%%%%%%%%%%%%%%%%%%%%%%%%%
%%%%%%%%%%%%%%%%%%%%%%%%%%%%%%%%%%%%%%%%%%%%%%%%%%%
%%%%%%%%%%%%%%%%%%%%%%%%%%%%%%%%%%%%%%%%%%%%%%%%%%%
%%%%%%%%%%%%%%%%%%%%%%%%%%%%%%%%%%%%%%%%%%%%%%%%%%%
\bigskip\bigskip\bigskip
\noindent
\underline{\bf Methods}

% \section{Dynamics of monitored JJA}\label{app:Dynamics}
\noindent {\bf Dynamics of monitored JJA}

Here we discuss the non-unitary time evolution of the JJA under weak measurements and feedback discussed in the main text [see Fig.\ref{fig1}(a)]. Before the $n$-th measurements at time $t=t_n^{-}=n\tau^{-}$, the whole system of JJA and the meters on the links $\langle ij\rangle$ are described by the density matrix $\rho_\mathrm{tot}(\{\xi_{ij}\},t_n^{-})=\otimes_{\langle ij\rangle}|\psi_{ij,n}\rangle\langle \psi_{ij,n}|\otimes U_s(\tau)\rho(\{\xi_{ij}\},t_{n-1}^{+}) U_s^\dagger(\tau)$, where the meters are prepared in a Gaussian states, $\psi_{ij}(x_{ij,n})=\langle x_{ij,n}|\psi_{ij}\rangle=(\pi\sigma)^{-1/4}\exp{(-x_{ij,n}^2/2\sigma)}$, and the JJA is in a state with (un-normalized) density matrix, $\rho(\{\xi_{ij}\},t_{n-1}^{+})$ at time $t_{n-1}$ after $(n-1)$-th set of measurements. The latter depends on all the previous measurement outcomes ${\xi_{ij}}$ till $t_{n-1}$, starting with some initial density matrix $\rho_0$ of the JJA. $U_s(\tau)=e^{-\mathrm{i}\mathcal{H}_s\tau/\hbar}$ is the unitary evolution due to JJA Hamiltonian $\mathcal{H}_s$. After the system-detector interaction $\mathcal{H}_{sd}(t)$ at time $t=n\tau$, the position of the meters are measured projectively at $t=n\tau^{+}$, leading to
\begin{align}
\rho_\mathrm{tot}(t_n^{+})&=\mathcal{P}_n[\prod_{\langle ij\rangle}(e^{-\mathrm{i}\Delta\hat{\theta}_{ij}\hat{p}_{ij,n}})\otimes_{ij}|\psi_{ij,n}\rangle\langle \psi_{ij,n}|\nonumber \\
&\otimes U_s(\tau)\rho(t_{n-1}^{+}) U_s^\dagger(\tau)\prod_{\langle ij\rangle }(e^{\mathrm{i}\Delta\hat{\theta}_{ij}\hat{p}_{ij,n}})]\mathcal{P}_n
\end{align}
where $\mathcal{P}_n=\otimes_{\langle ij\rangle}|\xi_{ij,n}\rangle \langle \xi_{ij,n}|$ is the projection operator corresponding to the measurement outcomes $\{\xi_{ij,n}\}$ in the $n$-th measurements. The density matrix of the JJA, $\rho(t_n^{+})=\mathrm{tr_M}[\rho_{tot}(t_n^{+})]=(\prod_{\langle ij\rangle } M_{ij,n})U_s(\tau)\rho(t_{n-1}^{+})U_s(\tau)(\prod_{\langle ij \rangle}M_{ij,n}) $, is obtained by tracing ($\mathrm{tr_M}$) over the meters, where $M_{ij,n}=\langle \xi_{ij,n}|e^{-\mathrm{i}\Delta\hat{\theta}_{ij}\hat{p}_{ij,n}}|\psi_{ij,n}\rangle=(\pi\sigma)^{-1/4}e^{-(\xi_{ij,n}-\Delta\hat{\theta}_{ij})^2/2\sigma}$ leads to non-unitary evolution of the state of the system due to measurements on the meters. As mentioned in the main text, we further apply a unitary displacement operator $U_{ij,n}=e^{\mathrm{i}\gamma \tau \xi_{ij,n}(\hat{n}_i-\hat{n}_j)}$ as a feedback \cite{Caves1987,Wiseman:2009rda} to control the heating of the JJA due to repeated measurements.

Accumulating all the steps of the above weak measurement setup, the density matrix of the JJA after $M$ measurements, over an interval $t_f=t_M^+$, is obtained as
\begin{align}
 \rho(\{\xi\},t_f)=[\prod_{n=1}^MK(\{\xi_{ij,n}\})U_s(\tau)]\rho_0 [\prod_{n=1}^M U_s^\dagger(\tau)K^\dagger(\{\xi_{ij,n}\})] \label{eq:Dynamics_rho_A} 
\end{align}
where  $K(\{\xi_{ij,n}\})=\prod_{\langle ij \rangle} U_{ij,n} \prod_{\langle ij \rangle} M_{ij,n}$ and $\rho_0$ is the initial density matrix at $t_0$. Finally, following procedure similar to that for monitored oscillators in Ref.\onlinecite{Ruidas2024} (see Supplementary Sec.\ref{sec:PathIntegral_S}), in the continuous weak measurement limit, $\tau\to 0,~\sigma\to \infty$ with $\sigma \tau=\Delta$ finite, we obtain the Schwinger-Keldysh generating function $Z[\xi]=\mathrm{Tr}[\rho(t_f,\{\xi\})]$ for a given quantum trajectory $\{\xi_{ij}(t)\}$ and the corresponding action of Eq.\eqref{eq:Action}.

The generating function $Z[\xi]$ gives the Born probability of the quantum trajectory $\{\xi_{ij}(t)\}$, and the expectation value of any operator $A$ in the long-time steady state is obtained as $\langle A\rangle_{\xi}=(1/Z[\xi])\int \mathcal{D}\theta A[\theta] e^{\mathrm{i}S[\theta,\xi]/\hbar}$. Further, the trajectory averaged expectation value, $\overline{\langle A\rangle}=(1/\mathcal{Z})\int \mathcal{D}\xi Z[\xi] \langle A\rangle_{\xi}$, can be obtained from the generating $\mathcal{Z}=\int \mathcal{D}\xi Z[\xi]$, which is equal to 1 due to normalization of the Born probability.\\\\

\noindent {\bf Self-consistent harmonic approximation (SCHA) and effective action} 
% \label{app:SCHA}

For the SCHA, we take a harmonic variational action
\begin{align}
&S_v[\theta,\xi]=\int_{t_0}^{t_f}dt[\sum_{i,s}s\{\frac{\hbar^2\dot{\theta}^2_{is}}{2E_c}+\frac{\hbar^2\gamma}{E_c}\dot{\theta}_{is}\sum_{j\in \mathrm{nn}_i} \xi_{ij}(t)\}\nonumber \\
&-\sum_{\langle ij\rangle,s}s\{\frac{1}{2}D_{ij,s}(t)\Delta\theta_{ij,s}^2-\frac{\mathrm{i}s\hbar}{2\Delta}\left(\xi_{ij}(t)-\Delta\theta_{ij,s})^2\right)\}], \label{eq:ActionVar}
\end{align}
where $D_{ij,s}(t)$ are the variational parameters and $\mathrm{nn}_i$ denotes the set of nearest neighbor sites of $i$. A variational principle can be obtained for the Born probability $Z[\xi]=e^{-F[\xi]}$ as 
\begin{align}
Z[\xi]=Z_v[\xi]\langle e^{\mathrm{i}(S-S_v)/\hbar}\rangle_v\geq Z_v[\xi]e^{\mathrm{i}\langle S-S_v\rangle_v/\hbar}\equiv e^{-F_v[\xi]},
\end{align}
leading to $F_v=-(\mathrm{i}/\hbar)\langle S-S_v\rangle_v-\ln{Z_v}\geq F$, where $\langle \cdots\rangle_v$ is expectation value with respect to the variational action. We obtain a self-consistency condition by minimizing $F_v$ with respect to $D_{ij,s}(t)$ (see Supplementary Sec.\ref{sec:SCHA_S}). Further, averaging over all the trajectories weighted with Born probability, and assuming a space-time translationally invariant steady state, i.e., $D_{ij,s}(t)=D$, we obtain the self-consistency condition of Eq.\eqref{eq:SelfConsistency}. The expectation value, $\overline{\langle \Delta\theta_{ij,s}^2(t)\rangle_v}=\mathrm{i}(1/Nd)\sum_q K(\mathbf{q})G^K(q)$, in Eq.\eqref{eq:SelfConsistency} is obtained using the trajectory averaged generating function $\mathcal{Z}=\int \mathcal{D}\theta e^{\mathrm{i}S_{eff}[\theta]/\hbar}$, with the effective action $S_{eff}$ of Eq.\eqref{eq:Seff} (see Supplementary Sec.\ref{sec:EffAction_S}).

% \begin{bibunit}[plain]
% \putbib[references]
% \end{bibunit}
 
\bibliographystyle{unsrtnat}
\bibliography{references} 

\clearpage
\newpage
%----------------------------------------------------------------
%--------------------------------------------------------------
\def\makeSM{1}
\ifdefined\makeSM

\appendix
\renewcommand{\appendixname}{}
\renewcommand{\thesection}{{S\arabic{section}}}
\renewcommand{\theequation}{\thesection.\arabic{equation}}
\renewcommand{\thefigure}{S\arabic{figure}}
\setcounter{page}{1}
\setcounter{figure}{0}
\setcounter{equation}{0}

\widetext

\centerline{\bf Supplemental Material}
\centerline{\bf for}
\begin{center}
\bf Inverse Superconductor-Insulator Transition in Weakly Monitored Josephson Junction Arrays
\end{center}
\centerline{ \authorf, \authors}
\centerline{\affiliationf}
\centerline{\affiliations}
\fi %%TO MAKE SUPPLEMENTAL MATERIAL

\def  \qinv{Q^{-1}}
\def  \q0{\frac{\w_k^2}{\G}+z}

\newcommand{\lin}{linspace}
\newcommand{\m}{\Delta_0}
\newcommand{\M}{\Delta}
\newcommand{\g}{Q}
\newcommand{\ginv}{\big(Q^{-1}\big)_{ab}}
\newcommand{\qr}{q_{\text{reg}}} 
\newcommand{\sr}{\Sigma_{\text{reg}}}
\newcommand{\qt}{\tilde{q}_{{EA}}}
\newcommand{\bfig}{\begin{figure}[H]\centering}
\newcommand{\efig}{\end{figure}}
\newcommand{\s}{\hspace{0.5cm }} % s for skip
\newcommand{\sh}{\hspace{0.25cm }} % sh for half skip.

\section{Schwinger-Keldysh (SK) path integral for monitored Josephson junction arrays (JJA)}\label{sec:PathIntegral_S}
In this section, we briefly discuss the path integral representations of $Z[\xi]=\mathrm{Tr}[\rho(\{\xi\},t_f)]$ for the density matrix evolution of Eq.\eqref{eq:Dynamics_rho_A} in the continuous weak measurement limit $\sigma\to\infty,~\tau\to 0$ with $\sigma\tau=\Delta$ finite. In this limit, we get 
\begin{equation}
\begin{split}
Z[\xi] 
 = \int_{\theta_{\pm}(t_f) = \theta (t_f)}    &{\displaystyle \prod_{i}} d\theta_i(t_f)  {\displaystyle \prod_{s = \pm 1,n=0}^{M-1}}d\theta_{is} (t_0+n\tau) \left< \theta_{+}(t_0)|\rho_0|\theta_{-}(t_0)\right> \\
 &  {\displaystyle \prod_{t = t_0}^{t_f -\tau}} \left< \theta_+(t+\tau)|e^{\mathrm{i}\gamma\tau\sum_{\left<ij\right>}\xi_{ij}(t+\tau)(\hat{n}_i-\hat{n}_j)}e^{-\tau\sum_{\left<ij\right>} [\xi_{ij} (t+\tau)-(\hat{\theta}_i-\hat{\theta}_j)]^2/2\Delta}e^{-\mathrm{i}\mathcal{H}_s\tau/\hbar}|\theta_+(t)\right>\\
 &  {\displaystyle \prod_{t = t_0}^{t_f -\tau}} \left< \theta_-(t+\tau)\right|e^{\mathrm{i}\gamma\tau\sum_{\left<ij\right>}\xi_{ij}(t+\tau)(\hat{n}_i-\hat{n}_j)}e^{-\tau\sum_{\left<ij\right>} [\xi_{ij} (t+\tau)-(\hat{\theta}_i-\hat{\theta}_j)]^2/2\Delta}e^{-\mathrm{i}\mathcal{H}_s\tau/\hbar}\left|\theta_-(t)\right>^*,\\
\end{split}
\label{equ:B1}
\end{equation}
where the matrix elements above can be evaluated as,
\begin{equation}
\begin{split}
 &\left< \theta(t+\tau)\right|e^{\mathrm{i}\gamma\tau\sum_{\left<ij\right>}\xi_{ij}(t+\tau)(\hat{n}_i-\hat{n}_j)}e^{-\tau\sum_{\left<ij\right>} [\xi_{ij} (t+\tau)-(\hat{\theta}_i-\hat{\theta}_j)]^2/2\Delta}e^{-\mathrm{i}\mathcal{H}_s\tau/\hbar}\left|\theta(t)\right>\\
  \sim &e^{\frac{\mathrm{i}\hbar}{2E_c\tau}\sum_i\left[\theta_i(t+\tau)-\theta_i(t)+\gamma\tau\tilde{\xi}_i(t+\tau)\right]^2}e^{-\frac{\mathrm{i}\tau}{\hbar}\sum_{\left<ij\right>}J_{ij}\left[1-\text{cos}(\theta_i(t)-\theta_j(t))\right]}e^{-\frac{\tau}{2\Delta}\sum_{\left<ij\right>}\left[\xi_{ij}(t+\tau)-\{\theta_i(t+\tau)-\theta_j(t+\tau)-\gamma\tau(\tilde{\xi}_i(t+\tau)-\tilde{\xi}_j(t+\tau))\}\right]^2} \\
\end{split}
\label{equ:B2}
\end{equation}
where $\tilde{\xi}_i = \sum_{j\in \text{nn}_i} \xi_{ij}$ with $\xi_{ji}=-\xi_{ij}$.
In the above, we do not explicitly include the pre-factor as it cancels between numerator and denominator while calculating expectation value of any observable using the generating function $Z[\xi]$. In the continuum limit, $\tau\to 0$, we obtain 
\begin{equation}
\begin{split}
Z[\xi] 
  &= \int_{\theta_{\pm}(t_f) = \theta (t_f)}    {\displaystyle \prod_{i}} d\theta_i(t_f)  {\displaystyle \prod_{s,n=0}^{M-1}}d\theta_{is} (t_0+n\tau) \left< \theta_{+}(t_0)|\rho_0|\theta_{-}(t_0)\right> \\
 & {\displaystyle \prod_{t = t_0,s}^{t_f -\tau}} e^{\frac{\mathrm{i}\hbar\tau}{2E_c}\sum_{i,s}s\left[\dot{\theta}_{is}(t)^2+2\dot{\theta}_{is}(t)\gamma\tilde{\xi}_i(t^+)\right]-\frac{\mathrm{i}\tau}{\hbar}\sum_{\left<ij\right>,s}sJ_{ij}[1-\text{cos}(\theta_{is}(t)-\theta_{js}(t))]}e^{-\frac{\tau}{2\Delta}\sum_{\left<ij\right>}\left[\xi_{ij}(t^+)-(\theta_{is}(t^+)-\theta_{is}(t^+))\right]^2} \\
 &=\int \mathcal{D}\theta e^{\mathrm{i}S[\theta,\xi]/\hbar}\left< \theta_{+}(t_0)|\rho_0|\theta_{-}(t_0)\right>
\end{split}
\label{equ:B3}
\end{equation}
The above is a path integral on the SK contour of Fig.\ref{fig2}(a) with the action in Eq.\eqref{eq:Action} (main text). Further, for long-time steady state, we take $t_0\to -\infty$ and $t_f\to\infty$.

\section{Quantum state diffusion and Lindblad dynamics with feedback for monitored JJA and Bose-Hubbard model} \label{sec:Lindblad_S}
In the work we have adopted a SK path integral approach to describe the dynamics of JJA under continuous weak-measurements and feedback. However, in this limit of $\tau\to0$, the dynamics of Eq.\eqref{eq:Dynamics_rho_A} can be equivalently described in terms of a quantum state diffusion (QSD) or stochastic Schr\"{o}dinger equation \cite{Gisin1984_S,Ghirardi1990_S}. To this end, we consider the dynamics of a pure state $\rho_0=|\psi_0\rangle\langle\psi_0|$. In this case, the dynamics of Eq.\eqref{eq:Dynamics_rho_A} between $t=t_{n-1}^+$ and $t+\delta t=t_n^+$, with $\delta t=\tau$, can be written as,
\begin{align}
    |\psi(t+\delta t)\rangle&=e^{\mathrm{i}\gamma\sum_{\langle ij\rangle}\xi_{ij}(\hat{n}_{i}-\hat{n}_j)\delta t}e^{-\delta t\sum_{ij}(\xi_{ij}-\hat{\Delta\theta_{ij}})^2/2\Delta}e^{-\mathrm{i}\mathcal{H}_s\delta t}|\psi(t)\rangle,
\end{align}
where $\Delta\hat{\theta}_{ij}=\hat{\theta}_i-\hat{\theta}_j$, $\xi_{ij}\equiv \xi_{ij,n}$ and $|\psi(t)\rangle$ is a normalized wavefunction at time $t$. From the above, we can obtain \cite{Jacobs2006_S} the probability density $P(\{\xi_{ij}\})$ of the outcomes $\{\xi_{ij}\}$ at $t+\delta t$ as
\begin{align}
    \langle\psi(t+\delta t)|\psi(t+\delta t)\rangle\propto P(\{\xi_{ij}\})=\prod_{\langle ij\rangle}\left(\frac{\delta t}{\pi \Delta}\right)^{1/2}e^{-(\delta t/\Delta)(\xi_{ij}-\langle \Delta\hat{\theta}_{ij}(t)\rangle)^2}.
\end{align}
Thus, $\xi_{ij}$ is Gaussian random number with a mean $\langle\Delta \hat{\theta}_{ij}(t)\rangle=\langle \psi(t)|\Delta\hat{\theta}_{ij}|\psi(t)\rangle$ and standard deviation $(\Delta/2\delta t)^{1/2}$. Based on the above, we redefine $(\delta t/\Delta)(\xi_{ij}-\langle\Delta\hat{\theta}_{ij}(t)\rangle)\to \xi_{ij,t}$, and after some tedious but straightforward manipulations \cite{Jacobs2006_S}, we obtain the equation for the normalized state $|\psi(t+\delta t)\rangle/\langle\psi(t+\delta t)|\psi(t+\delta t)\rangle^{1/2}\to|\psi(t+\delta t)\rangle$ up to $\mathcal{O}(\delta t)$ as
\begin{align}
|\psi(t+\delta t)\rangle=\left[1-\mathrm{i}\tilde{\mathcal{H}}_s\delta t+\sum_{\langle ij\rangle}\xi_{ij,t}\left(\hat{L}_{ij}-\langle \Delta \hat{\theta}_{ij}\rangle\right)-\frac{\delta t}{4\Delta}\sum_{\langle ij\rangle} \left[\left(\hat{L}^\dagger_{ij}-\langle \Delta \hat{\theta}_{ij}\rangle\right)\hat{L}_{ij}-\left(\hat{L}_{ij}-\langle \Delta \hat{\theta}_{ij}\rangle\right)\langle \Delta \hat{\theta}_{ij}\rangle\right]\right]|\psi(t)\rangle. \label{eq:QSD_JJA_S}
\end{align}
Here 
\begin{subequations}
\begin{align}
\hat{L}_{ij}&=\hat{\theta}_i-\hat{\theta}_j+\mathrm{i}\gamma \Delta (\hat{n}_i-\hat{n}_{j}) \label{eq:JJAMeasOp_S}\\
\tilde{\mathcal{H}}_s&=\mathcal{H}_s-\frac{\gamma}{4}\sum_{\langle ij\rangle} \left[(\hat{n}_i-\hat{n}_{j})\Delta\hat{\theta}_{ij}+\Delta\hat{\theta}_{ij}(\hat{n}_i-\hat{n}_{j})\right] \label{eq:EffH_QSD_S}\\
\mathcal{H}_s&=\frac{1}{2}\sum_i E_c \hat{n}_i^2+J\sum_{\langle ij\rangle}\left(1-\cos\Delta\hat{\theta}_{ij}\right) \label{eq:JJA_H_S}\\
\langle \xi_{ij,t}\rangle&=0~~~~~\langle \xi_{ij,t}\xi_{kl,t'}\rangle=\frac{1}{2\Delta}\delta t \delta_{t,t'}\delta_{\langle ij\rangle,\langle kl\rangle}
\end{align}
\end{subequations}
The Eq.\eqref{eq:QSD_JJA_S} is in the form of a general QSD \cite{Gisin1984_S,Ghirardi1990_S} with an effective non-Hermitian measurement operator $\hat{L}_{ij}$ which involves both measurement and feedback operators, $\Delta\hat{\theta}_{ij}$ and $\Delta\hat{n}_{ij}$, respectively. The feedback also modifies the effective Hamiltonian $\tilde{\mathcal{H}}$ in Eq.\eqref{eq:EffH_QSD_S} that enters in the unitary part of the QSD dynamics of Eq.\eqref{eq:QSD_JJA_S}. The stochasticity of the QSD dynamics originates from the Gaussian random number $\xi_{ij,t}$ which has zero mean and variance $(\delta t/2\Delta)$. From the above QSD dynamics, we can obtain the time evolution of the trajectory-averaged density matrix $\rho(t)=\overline{|\psi(t+\delta t)\rangle\langle \psi(t+\delta t)|}$, which is averaged over the Gaussian random numbers $\xi_{ij,t}$, leading to a Lindblad master equation
\begin{align}
\frac{d\rho}{dt}=-\mathrm{i}[\tilde{\mathcal{H}}_s,\rho]+\frac{1}{2\Delta}\sum_{\langle ij\rangle} \left(\hat{L}_{ij}\rho\hat{L}_{ij}^\dagger-\frac{1}{2}\{\hat{L}_{ij}^\dagger \hat{L}_{ij},\rho\}\right) \label{eq:Lindblad_S}
\end{align}

\subsection{Monitored Bose-Hubbard model}
The monitored dynamics described above can be easily extended to more general model, namely the Bose-Hubbard (BH) model \cite{Fazio2001_S,Duchon2013_S},
\begin{align}
\mathcal{H}_s=-t\sum_{\langle ij\rangle} (a_i^\dagger a_j+a_j^\dagger a_i)+\frac{U}{2}\sum_i \hat{n}_i^2,
\end{align}
where $\{a_i,a_i^\dagger\}$ are usual bosonic operators and $\hat{n}_i=a_i^\dagger a_i$ is the number operator with eigenvalues $n_i=0,1,\cdots$.
In the limit of large number of bosons per site on average, the BH model can be mapped to the model of JJA [Eq.\eqref{eq:JJA_H_S}] through the transformation, $a_i^\dagger=\sqrt{\hat{n}_i}e^{\mathrm{i}\hat{\theta}_i}$, $a_i=e^{-\mathrm{i}\hat{\theta}_i}\sqrt{\hat{n}_i}$ with $[e^{\pm \mathrm{i}\hat{\theta}_i},\hat{n}_i]=\mp e^{\pm\mathrm{i}\hat{\theta}_i}$ or $[\hat{\theta}_i,\hat{n}_i]=\mathrm{i}$. Subsequently, $\sqrt{\hat{n}_i}$ is replaced by $\sqrt{\langle \hat{n}\rangle}$ in the kinetic energy term of the BH model to obtain the Josephson coupling term of the JJA with $2\langle \hat{n}\rangle t\to J$, where $\langle \hat{n}\rangle$ is the average occupation of boson. Furthermore, we redefine the boson number operator $\hat{n}_i-\langle \hat{n}\rangle \to \hat{n}_i$ in the interaction term of the BH model to obtain the charging term of the JJA, modulo a constant, assuming a homogeneous steady state with uniform $\langle \hat{n}_i\rangle=\langle \hat{n}\rangle$. The weak measurement of $\Delta\hat{\theta}_{ij}$ in the JJA at the link $\langle ij\rangle$ can be mapped to measurement of particle current in the BH model as, $2 \langle \hat{n}\rangle (\hat{\theta}_i-\hat{\theta}_j)\sim 2 \langle \hat{n}\rangle \sin(\hat{\theta}_i-\hat{\theta}_j)\sim \hat{J}_{ij}\equiv-\mathrm{i}(a_i^\dagger a_j-a_j^\dagger a_i)$. We apply the same feedback $U_{ij,n}=e^{\mathrm{i}\gamma \tau \xi_{ij,n}(\hat{n}_{i}-\hat{n}_j)}$. As a result, following the steps mentioned in the preceding section, the QSD dynamics for the BH model can be written as
\begin{align}
|\psi(t+\delta t)\rangle=\left[1-\mathrm{i}\tilde{\mathcal{H}}_s\delta t+\sum_{\langle ij\rangle}\xi_{ij,t}\left(\hat{\mathcal{L}}_{ij}-\langle \hat{J}_{ij}\rangle\right)-\frac{\delta t}{4\Delta}\sum_{\langle ij\rangle} \left[\left(\hat{\mathcal{L}}^\dagger_{ij}-\langle \hat{J}_{ij}\rangle\right)\hat{\mathcal{L}}_{ij}-\left(\hat{\mathcal{L}}_{ij}-\langle \hat{J}_{ij}\rangle\right)\langle \hat{J}_{ij}\rangle\right]\right]|\psi(t)\rangle, \label{eq:QSD_BH_S}
\end{align}
where 
\begin{subequations}
\begin{align}
\hat{\mathcal{L}}_{ij}&=\hat{J}_{ij}+\mathrm{i}\gamma \Delta (\hat{n}_i-\hat{n}_{j})\label{eq:JumpOp_S}\\
\tilde{\mathcal{H}}_s&=\mathcal{H}_s-\frac{\gamma}{4}\sum_{\langle ij\rangle} \left[(\hat{n}_i-\hat{n}_{j})\hat{J}_{ij}+\hat{J}_{ij}(\hat{n}_i-\hat{n}_{j})\right] \label{eq:EffH_BH_QSD_S}\\
\langle \xi_{ij,t}\rangle&=0~~~~~\langle \xi_{ij,t}\xi_{kl,t'}\rangle=\frac{1}{2\Delta}\delta t \delta_{t,t'}\delta_{\langle ij\rangle,\langle kl\rangle}
\end{align}
\end{subequations}
From the above, we obtain the Lindblad dynamics for the monitored BH model with feedback in the same form as in Eq.\eqref{eq:Lindblad_S}. The Lindblad jump operators of Eq.\eqref{eq:JumpOp_S} can be directly compared with the jump operator used for the dissipative current coupling in the work by Diehl et al.\cite{Diehl2008_S,Diehl2010b_S}, i.e., $\hat{\mathcal{L}}_{ij}=(a_i^\dagger+a_j^\dagger)(a_i-a_j)=-\mathrm{i}(\hat{J}_{ij}+\mathrm{i}(\hat{n}_i-\hat{n}_{j}))$, which maps to $\hat{\mathcal{L}}_{ij}\to 2\langle \hat{n}\rangle \hat{L}_{ij}$ in the JJA model with $\hat{L}_{ij}=\hat{\theta}_i-\hat{\theta}_j+\mathrm{i} (1/2\langle \hat{n}\rangle)(\hat{n}_i-\hat{n}_j)$ of Eq.\eqref{eq:JJAMeasOp_S} and  $\gamma\Delta=1/2\langle \hat{n}\rangle$. In addition, compared to refs.\onlinecite{Diehl2008_S,Diehl2010b_S}, there is an extra term in the unitary part of the Lindblad dynamics of our monitored JJA model due to the feedback as in Eq.\eqref{eq:EffH_BH_QSD_S}. The particular parameter choice, $\gamma\Delta=1/2\langle \hat{n}\rangle$ leads to $T_{eff}=E_c/(4\gamma \Delta)=E_c\langle \hat{n}\rangle/2$ in the monitored JJA model, in agreement with results for effective temperature of Diehl et al. \cite{Diehl2008_S,Diehl2010b_S}, who used a \emph{single-site} variational approximation where the density matrix is approximated in terms of product of density matrices for each momentum mode \cite{Diehl2008_S} or site \cite{Diehl2010b_S}. However, in our monitored system the independent tuning of the strength of the measurement $\Delta^{-1}$ and feedback $\gamma$ provide extra handles to
control both the jump operator [Eq.\eqref{eq:JumpOp_S}] and the effective Hamiltonian [Eq.\eqref{eq:EffH_BH_QSD_S}]. This allows us the explore new and much more extended regime of the dynamical phase diagram, not studied in refs.\onlinecite{Diehl2008_S,Diehl2010b_S}. Moreover, the SCHA approximation employed in our work, and the consideration of various asymptotically exact limiting cases (see main text), allow us to treat non-perturbative effect of quantum fluctuations beyond single-site mean field theory. Such treatments help us to unravel
unusual frequency-dependent steady-state fluctuation-dissipation relation (FDR) and reveal that a description solely in terms of an effective temperature $T_{eff}$, particularly when $T_{eff}\to 0$, is inadequate to describe the non-equilibrium steady state in general. The full frequency-dependent FDR gives rise to phenomenon like re-entrant superconductor-insulator transitions, without any analogue in thermal equilibrium.

\section{Variational principle and self-consistent harmonic approximation (SCHA) for the dynamics of monitored JJA}\label{sec:SCHA_S}

Here we discuss a variational principle for the Born probability $Z[\xi]=e^{-F[\xi]}$ of a quantum trajectory and its application in terms of a SCHA, as mentioned in the main text and in Methods. The Born probability $Z[\xi] = \int \mathcal{D}\theta e^{\mathrm{i}S\left[\theta,\xi\right]}$ can be expressed in terms of a variational action $S_v[\theta,\xi]$ and corresponding partition function $Z_v = \int \mathcal{D}\theta e^{\mathrm{i}S_v\left[\theta,\xi\right]}$ by the following expression,
\begin{align}
Z 
&= Z_v\left[\frac{1}{Z_v}\int \mathcal{D}\theta e^{\mathrm{i}(S-S_v)/\hbar}e^{\mathrm{i}S_v/\hbar}\right]= Z_v\left<e^{\mathrm{i}(S-S_v)/\hbar}\right>_v \label{equ:C1}
\end{align}
Using the identity $\left<e^A\right> \geq e^{\left<A\right>}$ we can write  $Z \geq Z_v e^{\mathrm{i}\left<S-S_v\right>_v/\hbar} $. As a result,
\begin{equation}
\begin{split}
 F[\xi] &\leq  -\text{log} Z_v- \mathrm{i}\left<S-S_v\right>_v/\hbar = F_v
\end{split} 
\label{equ:C2}
\end{equation}
To obtain self-consistent equation, we minimize the $F_v[\xi]$ with respect to the variational parameter $D_{ij,s}(t)$ in the variational action of Eq.\eqref{eq:ActionVar} and get
\begin{align}
&\int_{-\infty}^{\infty} dt \sum_{\langle ij\rangle,s} s[J\frac{\partial}{\partial D_{ij,s}(t)}\left[1-\left<\text{cos}(\theta_{is}(t)-\theta_{js}(t))\right>_v\right]-\frac{D_{ij,s}(t)}{2}\frac{\partial}{\partial D_{ij,s}(t)}\left<(\theta_{is}(t)-\theta_{js}(t))^2\right>_v]=0 
\end{align}
To further solve the above equation, we perform a trajectory averaging, assuming a long-time steady state with space-time translation invariance, i.e., $D_{ij,s}(t)=D$, leading to 
\begin{align}
J\frac{\partial}{\partial D}\overline{\langle \cos{(\Delta\theta_{ij,s}(t))}\rangle_v}+\frac{D}{2}\frac{\partial}{\partial D}\overline{\langle (\Delta\theta_{ij,s}(t))^2\rangle_v}=0. \label{eq:Minimization_S}
\end{align}
Where, 
\begin{align}
\overline{\left<\text{cos}[\Delta\theta_{ij,s}(t)]\right>_v }= \frac{1}{\mathcal{Z}}\mathrm{Re}\int\mathcal{D}\theta e^{\mathrm{i}S_{eff}[\theta]}e^{i(\theta_{is}(t)-\theta_{js}(t
))}  \label{eq:ExptCos_S}
\end{align}
for the effective action of Eq.\eqref{eq:Seff}. To this end,

\begin{align}
\mathrm{i}S_{eff}/\hbar + \mathrm{i}(\theta_{ks}(t) - \theta_{ls}(t))&= \frac{\mathrm{i}}{2\hbar}\int_{-\infty}^{\infty} dt  \sum_{ij} \vcenter{\hbox{$\begin{bmatrix}
    \theta_{ic}(t) & \theta_{iq}(t) \end{bmatrix}$}}\vcenter{\hbox{$\begin{bmatrix}
    G^{-1}_{ij,cc}(t,t') & G^{-1}_{ij,cq}(t,t')  \\
    G^{-1}_{ij,qc}(t,t')  & G^{-1}_{ij,qq}(t,t') \\
\end{bmatrix}$}} \vcenter{\hbox{$\begin{bmatrix}
    \theta_{ic}(t') \\
    \theta_{iq}(t') \\
\end{bmatrix}$}} \\
&+ \mathrm{i}  \sum_i\int_{-\infty}^{\infty} dt' \vcenter{\hbox{$\begin{bmatrix}
    \theta_{ic}(t) & \theta_{iq}(t) \end{bmatrix}$}}\vcenter{\hbox{$\begin{bmatrix}
    (\delta_{ik}-\delta_{il})\delta(t-t') \\
    s(\delta_{ik}-\delta_{il})\delta(t-t') \\
\end{bmatrix}$}},
\label{equ:C4}
\end{align} 
where we have performed the Keldysh rotation \cite{Kamenev2011_S} $\theta_{is}(t)=\theta_{ic}(t)+s\theta_{iq}(t)$, and the propagators $G^{-1}_{ij,\alpha\beta}$ ($\alpha,\beta=c,q$) are given in the main text below Eq.\eqref{eq:Seff}. Using the above and performing the Gaussian integrals in Eq.\eqref{eq:ExptCos_S}, we obtain

\begin{align}
\overline{\left<\text{cos}(\theta_{ks}(t)-\theta_{ls}(t
))\right>_{v}} &= \text{exp} \left(-\frac{i}{2}\sum_{ij} \vcenter{\hbox{$\begin{bmatrix}
     (\delta_{ik}-\delta_{il}) &  s(\delta_{jk}-\delta_{jl})\end{bmatrix}$}}\vcenter{\hbox{$\begin{bmatrix}
    G_{ij,cc} & G_{ij,cq}  \\
    G_{ij,qc}  & G_{ij,qq} \\
\end{bmatrix}$}}\vcenter{\hbox{$\begin{bmatrix}
     (\delta_{ik}-\delta_{il}) \\
     s(\delta_{ik}-\delta_{il})\\
\end{bmatrix}$}} \right)\\
&= \text{exp}\left(-\frac{1}{2} i\sum_{ij} (\delta_{ik}-\delta_{il})(\delta_{jk}-\delta_{jl})(G_{ij,cc}+G_{ij,cq}+G_{ij,qc}+G_{ij,qq})(t,t) \right)\\
\end{align}
It is straightforward to show that for the Gaussian action of Eq.\eqref{eq:Seff}, 
\begin{align}
\overline{\left<\theta_{ks}(t)-\theta_{ls}(t))^2\right>_{v}}=i\sum_{ij} (\delta_{ik}-\delta_{il})(\delta_{jk}-\delta_{jl})(G_{ij,cc}+G_{ij,cq}+G_{ij,qc}+G_{ij,qq})(t,t).
\end{align}
Hence, 
\begin{align}
\overline{\left<\text{cos}(\theta_{ks}(t)-\theta_{ls}(t
))\right>_{v}}&= \text{exp}\left(-\frac{1}{2} \overline{\left<\theta_{ks}(t)-\theta_{ls}(t))^2\right>_{v}}\right)
\label{equ:C5}
\end{align} 
Thus, we get
\begin{align}
\frac{\partial}{\partial D} \overline{\langle \cos{(\Delta\theta_{ij,s}(t))}\rangle_v}=\frac{1}{2}\overline{\langle \cos{(\Delta\theta_{ij,s}(t))}\rangle_v} \frac{\partial}{\partial D}\langle \overline{(\Delta\theta_{ij,s}(t))^2\rangle_v}
\end{align}
Using Eq.\eqref{eq:Minimization_S}, the above leads to the self-consistency condition of Eq.\eqref{eq:SelfConsistency} (main text).

\subsection{Effective action for SCHA}\label{sec:EffAction_S}
To obtain the effective action of Eq.\eqref{eq:Seff} for the trajectory averaged generating function $\mathcal{Z}=\int \mathcal{D}\theta e^{\mathrm{i}S_{eff}[\theta]}$, we first integrate out the measurement outcomes $\{\xi_{ij}(t)\}$,
\begin{align}
&\int \mathcal{D}\xi e^{-\int_{-\infty}^{\infty}dt [\frac{1}{\Delta}\sum_{\left<ij\right>}\xi_{ij}^2+ \frac{2\mathrm{i}}{\hbar} \sum_i \theta_{ic}\frac{\mathrm{i}\hbar}{\Delta}\tilde{\xi}_i+\frac{2\mathrm{i}}{\hbar}\sum_i\theta_{iq}m\gamma\partial_t\tilde{\xi}_i]}
\nonumber\\
&=\text{exp}\left\{\frac{\mathrm{i}}{\hbar}\int_{-\infty}^{\infty}dt\sum_{ij}\vcenter{\hbox{$\begin{bmatrix}
    \theta_{ic}(t) & \theta_{iq}(t) \\
\end{bmatrix}$}} \vcenter{\hbox{$\begin{bmatrix}
    -\mathrm{i}\frac{\hbar}{\Delta}K_{ij} & -\mathrm{i}\frac{\hbar^2\gamma}{E_c} K_{ij}(\mathrm{i}\partial_t-\mathrm{i}\eta)\\
    \mathrm{i}\frac{\hbar^2\gamma}{E_c} K_{ij}(\mathrm{i}\partial_t+\mathrm{i}\eta) & -\mathrm{i}\frac{\hbar^3\gamma^2}{E_c^2\Delta^{-1}}K_{ij}\partial_t^2\\
\end{bmatrix}$}}\vcenter{\hbox{$\begin{bmatrix}
    \theta_{jc} (t)\\
    \theta_{jq} (t)\\ 
\end{bmatrix}$}}\right\}
\label{equ:D1}
\end{align}
The above leads to the effective Gaussian action for SCHA,
\begin{align}
&S_{eff}= \int_{-\infty}^{\infty}dt\sum_{ij}\left[\begin{array}{cc}
\theta_{ic}(t) & \theta_{iq}(t)\end{array}\right]\nonumber\\
 ~~&\left[\begin{array}{cc}
0 & \frac{\hbar^2}{E_c}(\mathrm{i}\partial_{t}-\mathrm{i}\eta)^{2}\delta_{ij}-[D_{ij,c}(t)+{\mathrm{i}\frac{\hbar^2\gamma}{E_c}}(\mathrm{i}\partial_{t}-\mathrm{i}\eta)]K_{ij}\\
\frac{\hbar^2}{E_c}(\mathrm{i}\partial_{t}+\mathrm{i}\eta)^{2}\delta_{ij}-
[D_{ij,c}(t)-\mathrm{i}\frac{\hbar^2\gamma}{E_c}(\mathrm{i}\partial_{t}+\mathrm{i}\eta)]
K_{ij} & \mathrm{i}K_{ij}(\frac{\hbar}{\Delta}-{\frac{\hbar^3\gamma^{2}}{E_c^2\Delta^{-1}}}\partial_{t}^{2})
\end{array}\right] \left[\begin{array}{c}
\theta_{ic}(t)\\
\theta_{iq}(t)
\end{array}\right] \label{eq:SCHAEffAction_S}
\end{align}
For the space-time translationally invariant steady state, assuming $D_{ij,+}(t)=D_{ij,-}(t)=D$, and Fourier transforming to momentum and frequency space, we obtain the action of Eq.\eqref{eq:Seff} (main text).

By inverting the inverse propagator appearing in $S_{eff}$ of Eq.\eqref{eq:Seff}, we obtain the SCHA propagator which has the usual causal structure \cite{Kamenev2011_S},
\begin{align}
&G(\boldsymbol{q},\omega) =\vcenter{\hbox{$\begin{bmatrix}
    G^K(\boldsymbol{q},\omega) & G^R(\boldsymbol{q},\omega)\\
    G^A(\boldsymbol{q},\omega) & 0\\
\end{bmatrix}$}} \\
&G^R(\boldsymbol{q},\omega) = \frac{\hbar}{2}\frac{1}{\hbar^2(\omega+i\eta)^2/E_c - [D-\mathrm{i}\hbar^2\gamma(\omega+i\eta)/E_c]K(\boldsymbol{q})}\\
&G^A(\boldsymbol{q},\omega) = \frac{\hbar}{2}\frac{1}{\hbar^2(\omega-i\eta)^2/E_c - [D+\mathrm{i}\hbar^2\gamma(\omega-i\eta)/E_c]K(\boldsymbol{q})}\\
&G^K(\boldsymbol{q},\omega) =\frac{E_c^2}{2\hbar^3}\left[-iK(\boldsymbol{q})\left(\frac{\hbar}{\Delta}+\frac{\hbar^3\gamma^2}{E_c^2\Delta^{-1}}\omega^2\right)\right]\frac{1}{(\omega^2-\omega_{\boldsymbol{q}}^2)^2+\gamma^2K(\boldsymbol{q})^2\omega^2}\\
\label{equ:E1}
\end{align}
Here $\omega_{\boldsymbol{q}}^2=(E_c/\hbar^2)DK(\boldsymbol{q})$. Using the above propagators we obtain 
\begin{align}
\overline{\left<\Delta\theta_{ij,s}^2(t)\right>_v} &= \frac{E_c^2}{2\hbar^3 Nd} \sum_{\bm{q}}\int_{-\infty}^{\infty} \frac{d\omega}{2\pi} K^2(\boldsymbol{q})\left(\frac{\hbar}{\Delta}+\frac{\hbar^3\gamma^2}{E_c^2\Delta^{-1}}\omega^2\right)\frac{1}{(\omega^2-\omega_{\boldsymbol{q}}^2)^2+\gamma^2K(\boldsymbol{q})^2\omega^2}\nonumber\\
&=\frac{E_c}{4dD}\frac{\Delta^{-1}}{\gamma}+\frac{\gamma}{2\Delta^{-1}}
\label{eq:DeltaTheta_S}
\end{align}
The above is used in Eq.\eqref{eq:SelfConsistency} (main text) to compute the superfluid stiffness $D$ self-consistently and to obtain the SCHA phase diagrams, discussed in the main text.

\begin{figure}[ht]
\includegraphics[width=0.5\textwidth]{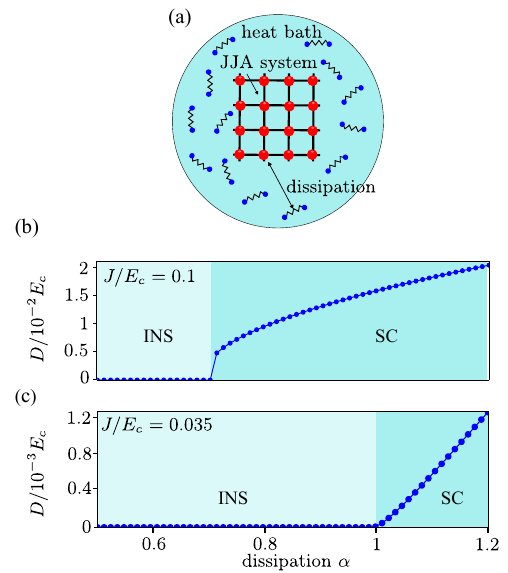}
    \caption{{\bf Ohmic JJA}:(a) The schematic diagram of a JJA coupled to an equilibrium Ohmic heat bath at constant temperatures $T$ is shown. Red circles denote the superconducting island and the black lines represent the Josephson junctions. Heat bath degrees of freedom are depicted as oscillators. (b) A horizontal cut through $J /E_c = 0.1$ is taken in Fig.\ref{fig1}(c) (main text) for $d=1$. We can see a jump in the value of $D$ at the first-order transition within SCHA. (c) A horizontal cut through $J /E_c = 0.035$ is taken in Fig.\ref{fig1}(d) (main text). We can see a continuous phase transition at $\alpha = 1/d$.}
    \label{fig:OhmicJJA_S}
\end{figure}

\section{Josephson junction arrays with Ohmic dissipation}~\label{sec:OhmicJJA_S}
In this section, we discuss the well-studied case \cite{Chak1986PRL_S,Chak1988PRB_S} of JJA in contact with an external equilibrium Ohmic bath at a fixed temperature $T$. 
The bath is modeled by harmonic oscillators \cite{Caldeira1981_S} as shown in Fig. \ref{fig2} (a), and the combined system of the JJA on a hypercubic lattice in $d$ dimension and the bath is described by the Hamiltonian,
\begin{subequations} \label{eq:HOhmicJJA_S}
\begin{align}
H &= H_s + H_b + H_{sb}\\
H_s &= \frac{1}{2}\sum_i E_c \hat{n}_i^2 + J\sum_{\left<ij\right>}(1-\text{cos}(\hat{\theta}_i - \hat{\theta}_j))\\
H_b &= \frac{1}{2}\sum_{\left<ij\right>}\sum_{l}  \left(\frac{\hat{p}_{l,ij}^2}{m_{l}}
+m_l^2\omega_{l}^2 \hat{x}^2_{l,ij}\right)\\
H_{sb} &= \sum_{\left<ij\right>}\Delta\hat{\theta}_{ij}\sum_{l}f_{l,ij}\hat{x}_{l,ij}
\label{equ:2}
\end{align}
\end{subequations}
where $f_{l,ij}$ are the coupling parameters between phase difference and the bath oscillators with position $\hat{x}_{l,ij}$, momenta $\hat{p}_{l,ij}$ and masses $m_l$. As in the model for monitored JJA, the dissipative coupling to the thermal bath here is through $\Delta\hat{\theta}_{ij}\sim \sin(\hat{\theta}_i-\hat{\theta}_j)$, or the current. 

 We employ the Schwinger-Keldysh (SK) path integral method \cite{Kamenev2011_S} for our analysis, as in the case monitored JJA. The SK generating function for the unitary evolution up to time $t_f$ of the combined JJA and bath system starting from a density matrix $\rho(t_0)$ at time $t_0$ is written as a path integral
 \begin{align}
Z(t_f)=\mathrm{Tr}[\rho(t_f)]=\mathrm{Tr}[\mathcal{U}(t_f,t_0)\rho(t_0)\mathcal{U}(t_0,t_f)]=\int \mathcal{D}x\mathcal{D}\theta e^{\mathrm{i}(S_s[\theta]+S_b[x]+S_{sb}[\theta,x]])/\hbar},
 \end{align}
 where $\mathcal{U}(t_f,t_0)=e^{-\mathrm{i}H(t_f-t_0)/\hbar}$. It is straightforward to obtain the actions $S_s[\theta],~S_b[x]$ and $S_{sb}[\theta,x]$ from the Hamiltonian of Eqs.\eqref{eq:HOhmicJJA_S}. We further integrate out the harmonic bath variables to obtain $Z=\int \mathcal{D}\theta e^{\mathrm{i}S[\theta]/\hbar}$ with an effective action $S[\theta]$ and assume usual Ohmic bath spectral function \cite{Kamenev2011_S,Fazio2001_S,Chak1986PRL_S,Chak1988PRB_S}.

\subsection{Self-consistent harmonic approximation for Ohmic JJA}\label{sec:OhmicJJA_SCHA_S}
Following Refs.\onlinecite{Chak1986PRL_S,Chak1988PRB_S}, and as done for monitored JJA, we apply SCHA, where a variational action $S_v[\theta]$ is written by replacing $ J(1-\text{cos}(\theta_i - \theta_j)) $ in $S[\theta]$ by the harmonic term
$(1/2)D(\theta_i - \theta_j)^2
$
where $D_{ij}$ is the variational parameter. 
Following procedure similar to the case of monitored JJA in Sec.\ref{sec:SCHA_S}, we get the following self-consistency condition \cite{Chak1986PRL_S},
\begin{equation}
\begin{split}
D =  J\text{exp}\left(-\frac{1}{2}\left<(\theta_i-\theta_j)^2\right>_{v}\right). 
\end{split}
\label{equ:3}
\end{equation}
Within SCHA, the action can be written as
\begin{align}
\frac{1}{\hbar}S_v=\frac{1}{2}\sum_q \theta^T(-q)G^{-1}(q)\theta(q),
\end{align}
in terms of Fourier transform $\theta^T(q)=(\phi_c(q),\phi_q(q))$ of the classical and quantum components of phase fields $\theta_{is}(t)=\theta_{ic}(t)+s\theta_{iq}(t)$, with $q=(\boldsymbol{q},\omega)$ and $\sum_q=\sum_{\boldsymbol{q}}\int(d\omega/2\pi)$. The inverse propagators are 
\begin{subequations}
\begin{align}
[G^{-1}]^{R(A)}(q)&=\frac{2}{\hbar}\left[\frac{\hbar^2}{E_c}\left((\omega\pm \mathrm{i}\eta)^2-\omega_{\boldsymbol{q}}^2\right)\pm\frac{\mathrm{i}\alpha\hbar\omega}{2\pi}K(\boldsymbol{q})\right]\\
[G^{-1}]^K(q)&=\cot{\left(\frac{\hbar\omega}{2T}\right)}\left([G^{-1}]^R(q)-[G^{-1}]^A(q)\right),
\end{align}
\end{subequations}
Here we impose FDR on the Keldysh component for thermal equilibrium at temperature $T$ ($k_{\mathrm{B}}=1$).  
The dimensionless parameter $\alpha = h/4e^2R$, where $R$ is the normal state resistance, controls the strength of dissipation. Using the variational action, we obtain  
\begin{equation}
\begin{split}
\left<\Delta\theta_{ij}^2\right>_v =\frac{\hbar}{2Nd} &\sum_{\boldsymbol{q}}\int_{-\infty}^{\infty} \frac{d\omega}{2\pi} K(\boldsymbol{q})\text{coth}\left(\frac{\hbar \omega}{2T}\right)\left(\frac{2\hbar\omega\frac{\alpha}{2\pi} K(\boldsymbol{q})}{\frac{\hbar^4}{E_c^2}(\omega^2-\omega_{\boldsymbol{q}}^2)^2+\hbar^2\omega^2\frac{\alpha^2}{4\pi^2}K^2(\boldsymbol{q})}\right)\\
\end{split}
\label{equ:5}
\end{equation}
We reproduce the phase diagram of Refs.\onlinecite{Chak1986PRL_S,Chak1988PRB_S} for Ohmic JJA by solving Eq.\eqref{equ:3} self-consistently using Eq.\eqref{equ:5} within the Debye approximation $K(\bm{q}) \sim q^2$, as shown in Fig.\ref{fig1}(c,e) (main text). The line depicts the phase boundary between the insulating normal state and the superconducting state. The superfluid stiffness $D$ (color) changes discontinuously across the solid line, as shown in Fig.\ref{fig:OhmicJJA_S}(c) for $J/E_c=0.1$, whereas $D$ changes continuously across the dashed line, as shown in Fig.\ref{fig:OhmicJJA_S}(d) for $J/E_c=0.035$. The vertical boundary at $\alpha = 1/d$ is a
line of continuous transition. For $\alpha<1/d$ the quantum fluctuations destroy the long-range phase coherence
even at $T =0$ for $J/E_c$ below a critical value that varies with $\alpha$. From Fig.\ref{fig1}(c) (main text), it is evident that for $J/E_c$ values below a threshold, the phase transition is insensitive to $J/E_c$, depending only on $\alpha$. Thus, the transition is solely governed by dissipation parameter $\alpha$ and beyond the critical value $\alpha = 1/d$, quantum fluctuations of Cooper pair number are entirely suppressed by dissipation, always leading to a phase coherent superconducting state at $T=0$. We show the phase diagram in $T-J$ plane in Fig.\ref{fig1} (e) (main text) for a fixed $\alpha = 0.5$. Here, the critical value of $J/E_c$ increases with temperature, as expected due to the increase of thermal fluctuations with temperature. Qualitatively similar phase diagrams are obtained within SCHA for higher dimensions \cite{Chak1986PRL_S,Chak1988PRB_S}.

We now compare the Ohmic JJA at high temperatures with monitored JJA within SCHA. At high temperature $T$, we expand $\coth{(\hbar\omega/2T)}$ appearing in Eq.\eqref{equ:5} up to $\mathcal{O}\left(\hbar\omega/T\right)$ to get,
\begin{equation}
\begin{split}
\left<\Delta\theta_{ij}^2\right>_{\text{var}}=\frac{1}{2Nd} &\sum_{\boldsymbol{q}}\int_{-\infty}^{\infty} \frac{\hbar d\omega}{2\pi}
\left(1+\frac{\hbar^2\omega^2}{12T^2}\right)\left(\frac{4T\frac{\alpha}{2\pi} K^2(\boldsymbol{q})}{\frac{\hbar^4}{E_c^2}(\omega^2-\omega_{\boldsymbol{q}}^2)^2+\hbar^2\omega^2\frac{\alpha^2}{4\pi^2}K^2(\boldsymbol{q})}\right)
\end{split}
\label{equ:15}
\end{equation}
By comparing Eq.\eqref{eq:DeltaTheta_S} and Eq.\eqref{equ:15}, we see the similarities between Ohmic and monitored JJAs, namely $\hbar\gamma/E_c \sim \alpha/2\pi$ and $\hbar\Delta^{-1}/E_c \sim 4 \gamma T$. Hence, the feedback strength of monitored JJA $\gamma$ is equivalent to dissipation strength $\alpha$ in the Ohmic JJA within SCHA. Additionally, the measurement strength $\Delta^{-1}$ exhibits behavior akin to the effective temperature $T_{eff}$ (see main text) when $\gamma$ is held constant, resembling the equilibrium temperature $T$ of the Ohmic JJA.

\section{Discussion on the phase diagram of monitored JJA} \label{sec:PhaseDiagramAnalytic_S}
We show the phase diagram of the monitored JJA in Figs.\ref{fig1}(b,d), in the $J-\gamma$ and $T_{eff}-J$ planes, and in Fig.\ref{fig2}(b) for $J-\Delta^{-1}$. The phase diagrams are numerically obtained from the self-consistent solution of Eq.\eqref{eq:SelfConsistency} (main text) for $d=1$. We observe that below a certain threshold value of $J/E_c$, the system remains insulating for any value of $\gamma, ~T_{eff}$ and $\Delta^{-1}$. Above the threshold $J/E_c$, the system undergoes a re-entrant insulator-superconductor-insulator transition as a function of $\gamma, ~T_{eff}$ and $\Delta^{-1}$. For the re-entrant phase transition observed in Fig.\ref{fig2}(b), there is a significant jump in the observed value of superconducting stiffness $D$ across the first transition and a smaller jump at the second transition (see Fig.\ref{fig2}(c,d)]).
 
We can gain analytical insights into the numerically obtained phase diagram. For any finite feedback $\gamma$, we get the self-consistent condition, $D/E_c = (J/E_c)\exp{( -\Delta^{-1}E_c/8d\gamma D-\gamma/4\Delta^{-1}})$. This condition can be expressed as,
$f(D) = \text{ln}\left(D/E_c\right) + \Delta^{-1}E_c/8d\gamma D=\text{ln}\left(J/E_c\right) -  \gamma/4\Delta^{-1}$.
For a fixed value of $\gamma$ and $\Delta^{-1}$, the minimum of $f(D)$ reveals the discontinuity in $D$ at the critical point, indicating a first-order phase transition within the SCHA, with the discontinuity given by,
\begin{equation}
\begin{split}
D_c/E_c = \frac{\Delta^{-1}}{8d\gamma}
\end{split}
\label{equ:16}
\end{equation}
This value is consistent with the results depicted in Fig.\ref{fig1}(b) and Fig.[\ref{fig2}(b)]. By substituting this value, we derive the equation for the phase boundary, 
\begin{equation}
\begin{split}
J_c/E_c = \frac{\Delta^{-1}}{8d\gamma} e^{1+\frac{\gamma}{4\Delta^{-1}}}
\end{split}
\label{equ:17}
\end{equation}
This equation also matches with the plots shown in Fig.\ref{fig1}(b) and Fig.\ref{fig2}(b). To find the threshold point, we minimize $J_c$, leading to
\begin{equation}
\begin{split}
\left(\frac{\gamma}{\Delta^{-1}}\right)_{th} = 4
\end{split}
\label{equ:18}
\end{equation}
This condition is also satisfied in Fig.\ref{fig1}(b) and Fig.\ref{fig2}(b). Substituting this in Eq.\eqref{equ:17}, we obtain the threshold value of coupling to be $J_{th} = \frac{e^2}{32d}$, which only depends on the dimension $d$. Thus, the threshold value gets lowered in higher dimensions.

\subsection{Comparison between dissipative JJA model and measurement JJA model }\label{sec:MonitoredOhmicComp_S}
We now contrast the monitored JJA with the Ohmic JJA. 
By comparing Eq.\eqref{eq:DeltaTheta_S} and Eq.\eqref{equ:5}, it is evident that a zero effective temperature is not achievable in the monitored model, unlike in Ohmic JJA model. In the zero-temperature limit of the Ohmic model, the integrand of Eq.\eqref{equ:5} will depend linearly on $\omega$. However, the measurement model has quadratic dependence without any linear term in $\omega$. 
Additionally, the limit $\Delta^{-1} \to 0$ while keeping $\gamma^2/\Delta^{-1}$ a constant value in the monitored model has no analogous limiting case possible for the Ohmic JJA. The integrand in Eq.\eqref{eq:DeltaTheta_S} for the monitored model, only contains the term of the order $\omega^2$, which is unattainable by an expansion of $\omega\hspace{0.1cm}\text{coth}\left(\omega/2T\right)$ (see Eq.\eqref{equ:5}) at any limit of $\omega$ or $T$ in the dissipative model. However, in that limit, the phase fluctuation of the measurement model in Eq.\ref{eq:DeltaTheta_S} diverges, always resulting in an insulating state.
 
 We now investigate some more interesting similarities and differences between the two models. The high-temperature Ohmic model and the monitored model, at high effective temperature $T_{eff}$, exhibit similar behavior, as discussed in main text. In Figs.\eqref{fig1}(d,e), we compare the phase diagram of dissipative model in $T-J$ plane with the phase diagram of the monitored model in $T_{eff}-J$ plane. 
We observe that the phase diagram of the Ohmic model at higher temperatures closely resembles the monitored model at the same effective temperature. However, at lower temperatures, their behavior differs significantly. One can observe that the critical value of the coupling strength increases with the equilibrium temperature in the Ohmic JJA. However, in the monitored JJA, we see the re-entrant phase transition while varying the effective temperature. Beyond a certain threshold value of $J/E_c$ we get an insulator-superconductor-insulator re-entrant phase transitions. At lower values of $T_{eff}$, we observe no global phase coherence. However, as the effective temperature rises above a critical value, it suppresses the phase fluctuations and restores superconductivity. Although at much higher effective temperatures, the model again fails to sustain the global phase coherence beyond a certain $T_{eff}$.

\section{Limiting cases of monitored JJA model}\label{sec:limits}

In this section, we discuss the various limiting cases of monitored JJA model to confirm the qualitative aspects of the SCHA phase diagrams discussed in the main text. 

\subsection{Semi-classical limit of monitored JJA}\label{sec:SemiClassical_S}
We take the semi-classical limit of our model following procedure similar to that in Ref.\onlinecite{Ruidas2024_S} for monitored oscillator chains. In the action of Eq.\eqref{eq:Action} (main text), we define $\bar{\xi}_{ij} = \xi_{ij}-\Delta\theta_{ij,c}$ and scale $\theta_{iq}$ and $\bar{\xi}_{ij}$ as $(\theta_{iq}/\hbar,\bar{\xi}_{ij}/\hbar) \to (\theta_{iq},\bar{\xi}_{ij})$. Subsequently, we expand the action $S[\theta,\xi]$ in powers of $\theta_{iq}$, which is equivalent to expansion in powers of $\hbar$. To approach a non-trivial semiclassical limit with $\hbar\to 0$, we consider $\Delta/\hbar^2,~E_c/\hbar^2$ to be finite in the limit $\hbar \to 0$, i.e., $\Delta,~E_c \sim \hbar^2,~\gamma \sim \mathcal{O}(\hbar^0)$ so that
$\theta_{ic},\theta_{iq},\bar{\xi}_{ij} \sim \mathcal{O}(1)$. This ensures that both $\bar{\xi}_{ij}$ and $\Delta\theta_{ij,q}$ are Gaussian distributed with a width of $\mathcal{O}(1)$ in the semi-classical limit. This also implies that $\xi_{ij}(t)=\Delta\theta_{ij,c}(t)+\hbar\mathcal{O}(1)$. Thus, the measurement outcomes $\{\xi_{ij}(t)\}$ get pinned to $\{\Delta\theta_{ij,c}(t)\}$ in the classical limit $\hbar\to 0$ and only a single quantum trajectory contributes to time evolution of the system. 
% Since $\gamma \sim \sqrt{\hbar/\Delta} \sim \sqrt{1/\hbar}$. 
Keeping the leading terms of $\mathcal{O}(1)$, we get,
\begin{align}
\frac{S}{\hbar}
=&\int dt \left\{\sum_i\left[ -2\frac{\hbar^2}{E_c}\ddot{\theta}_{ic}\theta_{iq}- 2\frac{\hbar^2\gamma}{E_c} \sum_{j\in \mathrm{nn}_i}\theta_{iq}\Delta\dot{\theta}_{ij,c}- \sum_{j\in \mathrm{nn}_i} 2\Delta\theta_{ij,q}\frac{\partial V}{\partial \Delta \theta_{ij,c}}\right] + \frac{i\hbar^2}{\Delta}\sum_{<ij>} \left[\bar{\xi}_{ij}^2+(\Delta\theta_{ij,q})^2\right]\right\}
\label{equ:19}
\end{align}
where, $V(\theta_{is} -\theta_{js}) = J(1-\text{cos}(\theta_{is} -\theta_{js}))$. In the above, we have used integration by parts while performing the integration $\int dt 2(\hbar^2/E_c)\dot{\theta}_{ic}\dot{\theta}_{iq}$ which becomes $-\int dt 2(\hbar^2/E_c)\ddot{\theta}_{ic}\theta_{iq}$, ignoring the boundary term. Similar integration is performed for $\int dt \theta_{ic}\dot{\theta}_{iq}$ which becomes $-\int dt\theta_{iq}\dot{\theta}_{ic}$. Additionally, we have utilized $\sum_i \dot{\theta}_{iq}\sum_{j\in \mathrm{nn}_i}\xi_{ij} = \sum_{<ij>}\Delta\dot{\theta}_{ij}^q\xi_{ij}$.

Using the Hubbard-Stratonovich identity we get 
\begin{align}
&Z[\xi]
\propto \int \mathcal{D}\eta \mathcal{D}\xi e^{-\frac{\hbar^2}{\Delta}\int dt \sum_{\langle ij\rangle}\bar{\xi}_{ij}^2} e^{-\frac{\Delta}{\hbar^2}\int dt \sum_{\langle ij\rangle}\eta_{ij}^2}\int \mathcal{D}\theta_c  {\displaystyle \prod_{it}} \delta \left(\frac{\hbar^2}{E_c}\ddot{\theta}_{ic} +\sum_{j\in \text{nn}_i} \left(\frac{\hbar^2\gamma}{E_c} \Delta\dot{\theta}_{ij}^c + \frac{\partial V}{\partial \Delta \theta_{ij}^c} -\eta_{ij}\right)\right)
\label{equ:F2}
\end{align}
This leads to the classical stochastic Langevin equation for the classical component, 
\begin{align}
\frac{\hbar^2}{E_c}\frac{d^2\theta_{ic}}{dt^2} =  \sum_{j\in \text{nn}_i}\left[-\frac{\hbar^2\gamma}{E_c} \left(\dot{\theta}_{ic} -\dot{\theta}_{jc}\right) 
- J\text{sin}(\theta_{ic}-\theta_{jc}) + \eta_{ij}\right]
\label{equ:20}
\end{align}
with a noise $\eta_{ij}$, which originates from quantum fluctuations $\theta_{iq}$. The noise has a Gaussian distribution controlled
by the measurement strength $\hbar^2/2\Delta$, such that for $\left<ij\right>$,
\begin{equation}
\begin{split}
\left<\eta_{ij}(t)\eta_{kl}(t')\right> &= \frac{\hbar^2}{2\Delta} \delta_{\langle ij\rangle, \langle kl\rangle}\delta(t-t')\\
\end{split}
\label{equ:F3}
\end{equation}

Here, we get $T_{eff} = (\Delta^{-1}/4\gamma)E_c \sim \mathcal{O}(\hbar^0)$. In the semiclassical limit, from Eq.\eqref{eq:DeltaTheta_S}, we obtain $\overline{\left<\Delta\theta_{ij,s}^2\right>}_v\simeq T_{eff}/dD+\mathcal{O}(\hbar^2)$.
As a result, from Eq.\eqref{eq:SelfConsistency}(main text), we get 
\begin{equation}
\begin{split}
D \simeq J e^{-\frac{T_{\text{eff}}}{2d D}}
\end{split}
\label{equ:21}
\end{equation}
Now we compare the classical limit of measurement model with the dissipative JJA model. In the high-temperature limit of the dissipative JJA model, we consider the expression of the phase fluctuation neglecting $\mathcal{O}(1/T)$ terms and beyond in Eq.\eqref{equ:5}. We obtain the phase fluctuations to be $\left<\Delta\theta_{ij}^2\right>_{\text{var}} \simeq T/dD$. Thus, the self-consistent equation becomes,
$D \simeq J \exp{[-(T/2 d D)]}$.
This expression is the same as Eq.\eqref{equ:21} for the monitored model in the semi-classical limit at effective temperature $T_{\text{eff}}$. Hence in the classical limit, both the monitored and dissipation cases exhibit the same behavior.

\subsection{Strong measurement limit of monitored JJA} \label{sec:StrongMeasurement_S}
As discussed in the preceding section, since $T_{eff}\propto \Delta^{-1}$, the limit of strong measurement $\hbar\Delta^{-1}\gg \hbar \gamma,J,E_c$ can be also thought of as an effective \emph{semiclassical} limit without assuming $\hbar\to 0$. To see this clearly, we scale $tE_c/\hbar\to t$ and $\sqrt{\hbar\Delta^{-1}/E_c}(\xi_{ij},\Delta\theta_{iq})\to (\xi_{ij},\Delta\theta_{iq})$ (after $\xi_{ij}-\Delta\theta_{ij,c}\to \xi_{ij}$) in the action [Eq.\eqref{eq:Action}], to obtain\begin{align}
\frac{S}{\hbar}= & \int dt\{2\sqrt{\frac{E_{c}}{\hbar\Delta^{-1}}}\sum_{i}\left[-\theta_{iq}\ddot{\theta}_{ic}-\frac{\hbar\gamma}{E_{c}}\theta_{iq}\sum_{j\in nn_{i}}\Delta\dot{\theta}_{ij,c}+\frac{\hbar\gamma}{E_{c}}\sqrt{\frac{E_{c}}{\hbar\Delta^{-1}}}\dot{\theta}_{iq}\sum_{j\in nn_{i}}\xi_{ij}\right] \nonumber\\
 & -\frac{\hbar}{E_{c}}\sum_{\langle ij\rangle}\tilde{V}(\Delta\theta_{ij,c},\sqrt{\frac{E_{c}}{\hbar\Delta^{-1}}}\Delta\theta_{ij,q})+\mathrm{i}\sum_{\langle ij\rangle}\left(\xi_{ij}^{2}+\Delta\theta_{ij,q}^{2}\right)\}
\end{align}
where $\tilde{V}(\Delta\theta_c,\Delta\theta_q)=(1/2\hbar)\sum_s sV(\Delta\theta_c+s\Delta\theta_q)$ and $V(\theta)=J(1-\cos\theta)$. Expanding in powers of $\sqrt{E_c/\hbar\Delta^{-1}}\ll 1$ and keeping $\mathcal{O}(1)$ terms, we get
\begin{align}
\frac{S}{\hbar}= & \int dt\left\{2\sqrt{\frac{E_{c}}{\hbar\Delta^{-1}}}\sum_{i}\left[-\theta_{iq}\ddot{\theta}_{ic}-\frac{\hbar\gamma}{E_{c}}\theta_{iq}\sum_{j\in nn_{i}}\Delta\dot{\theta}_{ij,c}-\theta_{iq}\sum_{j\in nn_{i}}\frac{1}{E_{c}}\frac{\partial V}{\partial\Delta\theta_{ij,c}}\right]+\mathrm{i}\sum_{\langle ij\rangle}\left(\xi_{ij}^{2}+\Delta\theta_{ij,q}^{2}\right)\right\}
\end{align}
Introducing Hubbard-Stratonovich field, integrating out the quantum component of the field, as in Eq.\eqref{equ:F2}, and after restoring the unit of time $t\to tE_c/\hbar$, we again obtain the stochastic Langevin dynamics of Eq.\eqref{equ:20}. Thus, in the strong measurement limit ($\Delta\to 0$), the system becomes effectively classical with quantum trajectory determined by the stochastically evolving classical component, i.e., $\xi_{ij}(t)=\Delta\theta_{ij,c}+\Delta \mathcal{O}(1)$.

\subsection{Strong feedback and weak coupling limits of monitored JJA}\label{sec:StrongFeedbackRG_S}

Here we consider the strong feedback limit, where $\hbar\gamma\gg J,E_c$, and the weak coupling limit $J\ll E_c$. To this end, we decompose the phases into slow [$\theta_{is}^<(t)$] and fast [$\theta_{is}^>(t)$] modes, i.e., $\theta_{is}(t)=\theta_{is}^>(t)+\theta_{is}^<(t)$, with
\begin{align}
\hspace{1cm} \theta_{is}^{>}(t) & =\int_{\tilde{\omega}_c\leq |\omega|\leq \omega_c} \frac{d\omega}{2\pi} e^{-\mathrm{i}\omega t}\theta_{is}(\omega),\\
\hspace{1cm}\theta_{is}^{<}(t) &=\int_{0\leq |\omega|\leq \tilde{\omega}_c} \frac{d\omega}{2\pi} e^{-\mathrm{i}\omega t}\theta_{is}(\omega),\\
\label{equ:22}
\end{align}
The lower and upper frequency cutoffs, $\tilde{\omega}_c$ and $\omega_c$, respectively, for the fast and slow modes are chosen appropriately depending on the limit, as discussed below. We integrate out the fast mode and obtain and effective dynamics for the slow modes with a renormalized JJ coupling. We also similarly split the field, $\xi_{ij}(t)=\xi_{ij}^<(t)+\xi_{ij}^>(t)$, corresponding to the measurement outcomes, into slow and fast components. Treating $J$ perturbatively up to $\mathcal{O}(J)$, we obtain the trajectory-averaged generating function
\begin{subequations}\label{eq:IntFastMode_S}
\begin{align}
   &\mathcal{Z}
\sim \int \mathcal{D}\xi^<\mathcal{D}\theta^< e^{\frac{i}{\hbar}S_0^<[\theta^<,\xi^<]} \left<e^{\frac{i}{\hbar}\sum_s s \int_{-\infty}^{\infty} dt \sum_{<ij>} J \text{cos}(\theta_{is}-\theta_{js})}\right>_>, \\  
&\approx \int \mathcal{D}\xi^<\mathcal{D}\theta^<  e^{\frac{i}{\hbar}S_0^<[\theta^<,\xi^<]} e^{\frac{i}{\hbar}\sum_s s \int_{-\infty}^{\infty} dt \sum_{<ij>} J_{eff} \text{cos}(\Delta\theta_{ij,s}^<)}, \label{eq:EffectiveAction_S}\\
J_{eff}&\approx J e^{-\frac{1}{2}\left<(\Delta\theta_{ij,s}^{>})^2\right>_>}
\label{equ:27} 
\end{align}
\end{subequations}
where the Gaussian action,
\begin{align}
S_0^f&=2\int dt\left\{\frac{\hbar^2}{E_c}\left[-\theta_{iq}^f\ddot{\theta}_{ic}^f+\gamma \dot{\theta}_{iq}^f\sum_{j\in \mathrm{nn}_i}\xi_{ij}^f-\gamma \theta_{iq}^f\sum_{j\in \mathrm{nn}_i}\Delta\dot{\theta}_{ij,c}^f\right]+\frac{\mathrm{i}\hbar}{\Delta}\sum_{\langle ij\rangle} \left[(\xi_{ij}^f)^2+(\Delta\theta_{ij,q}^f)^2\right]\right\}~~~~f=<,>.\\
\end{align}
The expectation $\langle\cdots\rangle_>$ in Eqs.\eqref{eq:IntFastMode_S} is with respect to the Gaussian action $S_0^>[\theta^>,\xi^>]$ of the fast modes, where
\begin{align}
\left<(\Delta\theta_{ij,s}^{>})^2\right>_> = \frac{\hbar}{2Nd} \sum_{\boldsymbol{q}} K^2(\boldsymbol{q}) \int_{\tilde{\omega}_c\leq |\omega|\leq \omega_c} \frac{d\omega}{2\pi} \frac{\hbar\Delta^{-1}+\frac{\hbar^3\gamma^2\omega^2}{E_c\Delta^{-1}}}{\hbar^4\omega^4/E_c^2+\hbar^4\gamma^2\omega^2K^2(\boldsymbol{q})/E_c^2}
\label{eq:PhaseCorrelator_S}
\end{align}

{\bf Strong feedback limit $\hbar\gamma\gg J,E_c$ --} In this case we set $\omega_c=\hbar\gamma$ and obtain the lower cutoff as $\tilde{\omega}_c=JE_c/\hbar^2\gamma$ by comparing the dissipative feedback term and the JJ term in the action of Eq.\eqref{eq:Action}, i.e., $J\sim (\hbar^2\gamma/E_c)\tilde{\omega}_c$. For this limit, using $\omega_c\gg \tilde{\omega}_c$, we get
\begin{align}
\langle(\Delta\theta_{ij,s}^{>})^{2}\rangle_{>}\approx & \frac{1}{Nd}\sum_{\boldsymbol{q}}\frac{1}{2\pi}\left[\frac{\Delta^{-1}E_{c}}{\gamma J}+\left(\frac{\gamma K(\boldsymbol{q})}{\Delta^{-1}}-\frac{\Delta^{-1}E_{c}^{2}}{\hbar^2\gamma^{3}K(\boldsymbol{q})}\right)\frac{\pi}{2}\right].
\end{align}
 Thus, for $\gamma\gg\Delta^{-1},J,E_{c}$
\begin{align}
\langle(\Delta\theta_{ij,c}^{>})^{2}\rangle_{>}\approx & \frac{\gamma}{2\Delta^{-1}},
\end{align}
where we have used $\sum_{\boldsymbol{q}} K(\boldsymbol{q})=2dN$, $\sum_{\boldsymbol{q}} = N$. In the limit, $\Delta^{-1}\gg\gamma\gg J,E_{c}$
\begin{align*}
\langle(\Delta\theta_{ij,c}^{>})^{2}\rangle_{>}\approx &  \frac{E_{c}\Delta^{-1}}{2\pi J \gamma d}
\end{align*}
 As a result,
 \begin{subequations}
\begin{align}
J_{eff}\approx & J\left(1-\frac{\gamma}{4\Delta^{-1}}\right)\hspace{1em}\hspace{1em}\hbar\gamma\gg \hbar\Delta^{-1},J,E_{c}\\
\approx & J\left(1-\frac{E_{c}\Delta^{-1}}{4\pi J \gamma d}\right)\hspace{1em}\hspace{1em}\hbar\Delta^{-1}\gg\hbar\gamma\gg J,E_{c}
\end{align}
\end{subequations}
Hence $J_{eff}$ becomes negative in both the above limits of strong feedback, indicating destruction of superconducting steady state.

{\bf Weak coupling renormalization group (RG) in the limit $J\ll E_c$ --} In this case, we choose the upper cutoff $\omega_c$ in Eq.\eqref{eq:PhaseCorrelator_S} depending on the largest energy scale among $E_c,~\hbar\Delta^{-1}$ and $\hbar\gamma$. The lower cutoff is $\tilde{\omega}_c=\omega_c/b$, where $b=e^\delta l>1$.
In the effective action for the slow modes in Eq.\eqref{eq:EffectiveAction_S}, we rescale $\left(\theta_{is}(\omega), \omega, t, E_c\right) \to \left(\theta_{is}(\omega)/b, \omega b, t/b, E_cb\right) $. This results in a similar expression for action with a rescaled JJ coupling by $J' = Jb e^{-\frac{1}{2}\left<(\Delta\theta_{ij,s}^>)^2\right>_>}$, where
\begin{align}
\left<(\Delta\theta_{ij,s}^{>})^2\right>_> = \frac{\hbar}{2Nd} \sum_{\boldsymbol{q}} \int_{\omega_c/b\leq |\omega|\leq \omega_c} \frac{d\omega}{2\pi} \frac{K^2(\boldsymbol{q})(\hbar\Delta^{-1}+\frac{\hbar^3\gamma^2\omega^2}{E_c\Delta^{-1}})}{\hbar^4\omega^4/E_c^2+\hbar^4\gamma^2\omega^2K^2(\boldsymbol{q})/E_c}
\label{equ:28}
\end{align}
Performing infinitesimal RG transformation $b = e^{\delta l} \approx  1 + \delta l$, we obtain the differential RG flow equation for $J$. For notational convenience, we define dimensionless quantities $\tilde{\gamma}=\hbar\gamma/E_c$ and $\tilde{\Delta}^{-1}=\hbar\Delta^{-1}/E_c$ below.

For $\hbar \gamma$ the largest energy scale, we take $\omega_c=\gamma$. Taking $\hbar\gamma$ much larger than other energy scales, we simplify the evaluation of $\langle(\Delta\theta_{ij,s}^>)^2 \rangle_>$ by ignoring the $\omega^4$ term in the denominator of Eq.\eqref{equ:28} to get $ \left<(\Delta\theta_{ij,s}^{>})^2\right>_> =\frac{1}{2\pi \tilde{\gamma} d} \left(\frac{\tilde{\Delta}^{-1}}{\tilde{\gamma}^2}+\frac{\tilde{\gamma}^2}{\tilde{\Delta}^{-1}}\right)\left(1-\frac{1}{b}\right)$. This gives the RG flow equation,
\begin{equation}
\begin{split}
\frac{dJ}{dl} &= J\left(1-\frac{1}{4\pi \tilde{\gamma} d} \left(\frac{\tilde{\Delta}^{-1}}{\tilde{\gamma}^2}+\frac{\tilde{\gamma}^2}{\tilde{\Delta}^{-1}}\right)\right),
\label{equ:29}
\end{split}
\end{equation}
leading to the $\beta$ function mentioned in the main text.
 In the case, where $E_c$ is the dominant energy scale, we cannot simply neglect the $\omega^4$ term in the denominator of the integrand in Eq.\eqref{equ:28}. However, we can still integrate out the modes between $\hbar \gamma<|\omega|<E_c$ to get an effective action whose upper frequency cutoff is determined by $\gamma$. This procedure merely multiplies a prefactor to the JJ term, keeping the same flow equation as above with an effective $J$. Hence we get the same RG flow equation Eq.\eqref{equ:29} also in this limit.

%%%%%%%%%%%%%%%%%%%%%%%%%%%%%%%%%%%%%%%%%%%%%%%%%%%
%%%%%%%%%%%%%%%%%%%%%%%%%%%%%%%%%%%%%%%%%%%%%%%%%%%
%%%%%%%%%%%%%%%%%%%%%%%%%%%%%%%%%%%%%%%%%%%%%%%%%%%
%%%%%%%%%%%%%%%%%%%%%%%%%%%%%%%%%%%%%%%%%%%%%%%%%%%
%%%%%%%%%%%%%%%%%%%%%%%%%%%%%%%%%%%%%%%%%%%%%%%%%%%

%%%%%%%%%%%%%%%%%%%%%%%%%%%%%%%%%%%%%%%%%%%%%%%%%%%
%%%%%%%%%%%%%%%%%%%%%%%%%%%%%%%%%%%%%%%%%%%%%%%%%%%
%%%%%%%%%%%%%%%%%%%%%%%%%%%%%%%%%%%%%%%%%%%%%%%%%%%
%%%%%%%%%%%%%%%%%%%%%%%%%%%%%%%%%%%%%%%%%%%%%%%%%%%
%%%%%%%%%%%%%%%%%%%%%%%%%%%%%%%%%%%%%%%%%%%%%%%%%%%

%%%%%%%%%%%%%%%%%%%%%%%%%%%%%%%%%%%%%%%%%%%%%%%%%%%
%%%%%%%%%%%%%%%%%%%%%%%%%%%%%%%%%%%%%%%%%%%%%%%%%%%
%%%%%%%%%%%%%%%%%%%%%%%%%%%%%%%%%%%%%%%%%%%%%%%%%%%
%%%%%%%%%%%%%%%%%%%%%%%%%%%%%%%%%%%%%%%%%%%%%%%%%%%
%%%%%%%%%%%%%%%%%%%%%%%%%%%%%%%%%%%%%%%%%%%%%%%%%%%

% \begin{bibunit}[unsrt]
% \putbib[references]
% \end{bibunit}

% \bibliographystyle{unsrtnat}
%\bibliography{supp_references} 

\end{document}